\documentclass[a4paper,fleqn,usenatbib]{mnras}

\usepackage{newtxtext,newtxmath}

\usepackage[T1]{fontenc}
\usepackage{ae,aecompl}

\usepackage{graphicx}	
\usepackage{subfigure}
\usepackage{amsmath}	
\usepackage{amssymb}	

\title[Sulphur chemistry in the L1544 pre-stellar core]{Sulphur chemistry in the L1544 pre-stellar core}

\author[C. Vastel et al.]{
Charlotte Vastel,$^{1}$\thanks{E-mail: charlotte.vastel@irap.omp.eu} 
D. Qu\'enard,$^{2}$ 
R. Le Gal,$^{3}$
V. Wakelam,$^{4}$
A. Andrianasolo,$^{5}$
\newauthor
P. Caselli,$^{6}$ 
T. Vidal,$^{4}$
C. Ceccarelli,$^{5}$
B. Lefloch,$^{5}$
and R. Bachiller$^{7}$
\\
$^{1}$IRAP, Universit\'e de Toulouse, CNRS, UPS, CNES, Toulouse, France\\
$^{2}$School of Physics and Astronomy, Queen Mary University of London, Mile End Road, London E1 4NS, UK\\
$^{3}$Harvard-Smithsonian Center for Astrophysics, 60 Garden Street, Cambridge, MA 02138, USA\\
$^{4}$Laboratoire d'astrophysique de Bordeaux, Univ. Bordeaux, CNRS, B18N, all\'ee Geoffroy Saint-Hilaire, Pessac, 33615, France\\
$^{5}$Univ. Grenoble Alpes, CNRS, IPAG, F-38000 Grenoble, France\\
$^{6}$Centre for Astrochemical Studies, Max-Planck-Institute for Extraterrestrial Physics, Giessenbachstrasse 1, 85748 Garching, Germany\\
$^{7}$Observatorio Astron\'omico Nacional (OAN, IGN), Calle Alfonso XII, 3, 28014 Madrid, Spain\\
}

\date{Accepted XXX. Received YYY; in original form ZZZ}

\pubyear{2018}

\begin{document}
\label{firstpage}
\pagerange{\pageref{firstpage}--\pageref{lastpage}}
\maketitle

\begin{abstract}
The L1544 pre-stellar core has been observed as part of the ASAI IRAM 30m Large Program as well as follow-up programs. These observations have revealed the chemical richness of the earliest phases of low-mass star-forming regions. In this paper we focus on the twenty-one sulphur bearing species (ions, isotopomers and deuteration) that have been detected in this spectral-survey through fifty one transitions: CS, CCS, C$_3$S, SO, SO$_2$, H$_2$CS, OCS, HSCN, NS, HCS$^+$, NS$^+$ and H$_2$S. We also report the tentative detection (4 $\sigma$ level) for methyl mercaptan (CH$_3$SH). LTE and non-LTE radiative transfer modelling have been performed and we used the \textsc{nautilus} chemical code updated with the most recent chemical network for sulphur to explain our observations. From the chemical modelling we expect a strong radial variation for the abundances of these species, which mostly are emitted in the external layer where non thermal desorption of other species has previously been observed. We show that the chemical study cannot be compared to what has been done for the TMC-1 dark cloud, where the abundance is supposed constant along the line of sight, and conclude that a strong sulphur depletion is necessary to fully reproduce our observations of the prototypical pre-stellar core L1544. 
\end{abstract}



\section{Introduction}
Sulphur is the tenth most abundant element in our Galaxy and has been the subject of a lot of controversy over the past 20 years. Most observations of diffuse media (probing different conditions) find no elemental depletion of sulphur \citep[e.g.][]{howk2006}. With a detailed analysis, \citet{jenkins2009} seems to show a depletion with increasing density (as for the other elements) but discussed the possible observational bias of this result. On the contrary, only a small fraction of the cosmic sulphur abundance is seen in cold dense cores (through observable molecules such as SO and CS) \citep[e.g.][]{tieftrunk1994,palumbo1997}. To reproduce these small abundances, chemical models need to deplete the elemental abundance of sulphur to "hide" the overflow of sulphur  \citep[see for example][]{wakelam2004}. The explanation commonly proposed is that the missing sulphur is locked on the icy mantles of dust grains \citep[e.g.][]{millar1990,ruffle1999}. The controversy resides in the fact that, until now, only OCS \citep{geballe1985,palumbo1995,palumbo1997} and possibly SO$_2$ \citep{boogert1997,zasowski2009} have been detected in solid state towards high-mass protostars, with abundances ($\sim$ 10$^{-7}$) less than 4\% of the sulphur cosmic abundance. H$_2$S, the most likely natural product of hydrogenation of sulphur on grains, has not been detected and the upper limits on the abundance are $3 \times 10^{-7}$ and $3\times 10^{-6}$ (assuming an abundance of H$_2$O in ices of $10^{-4}$) \citep{smith1991}. The long standing question is then where is the sulphur that appears to be depleted from the gas phase in the dense regions of the interstellar medium?\\
Experimental measurements have been performed in order to understand which species are good candidates for sulphur in its solid state. OCS, for example, is formed by cosmic-ray irradiation, although it is easily destroyed on a long term \citep{garozzo2010}. \citet{ferrante2008} discussed the alternative of carbon disulphide (CS$_2$), which has been detected in the coma of comet 67P/Churyumov-Gerasimenko \citep{calmonte2016}, although not detected in ices yet. Hydrated sulphuric acid (H$_2$SO$_4$) was also suggested by \citet{scappini2003} as the main reservoir. Photoproducts of H$_2$S ice processing were proposed as a plausible explanation of the absence of H$_2$S in the ices and the sulphur depletion towards dense clouds and protostars \citep{jimenez-escobar2011}. A large fraction of the missing sulphur in dense clouds could thus be polymeric sulphur residing in dust grains as proposed by \citet{wakelam2004}. Finally, \citet{druard2012} have proposed polysulphanes (H$_2$S$_n$) as possible carriers for sulphur on grains.\\

So far, only a few studies of sulphur chemistry have been performed in the cold regions of the interstellar medium. A sulphur depletion factor of $\sim$ 100 has been adopted to explain the chemistry in starless cores \citep{tafalla2006,agundez2013}. A higher gas-phase sulfur abundance approaching the cosmic value of 1.5 $\times$ 10$^{-5}$ has been found in bipolar outflows \citep{bachiller1997,anderson2013}, photodissociation regions \citep{goicoechea2006}, and hot cores \citep{esplugues2014}. In these cases, this abundance was possibly interpreted by the release of the sulphur bearing species from the icy grain mantles because of thermal and non-thermal desorption and sputtering. Very recently, such a study has been done by \citet{fuente2016} towards the Barnard B1b globule that hosts two candidates for the first hydrostatic core (FHSC). Their pointed position lies in between the two cores (B1b-N and B1b-S) leading to a difficult analysis. Their observational data are fitted using a chemical modelling with an elemental depletion of $\sim$ 25 for sulphur. \citet{fuente2016} proposed that the low sulfur depletion and high abundances of complex molecules could be the result of two factors, both related to the star formation activity: the enhanced UV fields and the surrounding outflows. The star formation activity could also have induced a rapid collapse of the B1b core that preserves the high abundances of the sulphured species. In addition, the outflow associated with B1b-S may heat the surroundings and contribute to stop the depletion of S-molecules. More recently, \citet{vidal2017} compared the observations of sulphur bearing species towards the TMC-1 (CP) dark cloud with the results from an updated chemical model and concluded that sulphur depletion is not required although a factor of three depletion also fitted the observations.

As part of the IRAM-30m Large Program ASAI\footnote{Astrochemical Surveys At Iram: http://www.oan.es/asai/} \citep{lefloch2018}, we carried out a highly sensitive, unbiased spectral survey of the molecular emission of the L1544 pre-stellar core with a high spectral resolution. In the present study we report on the detection of twenty-one sulphur bearing species in this core, use a radiative transfer modelling to determine the observed column densities and compare with the most up to date chemical modelling for sulphur chemistry.\\

We present in Section 2 the observations from the ASAI spectral survey and the line identification for the sulphur bearing species. Based on the detections and tentative detection, we compute in Section 3 the column densities of these species. We present in Section 4 the deuterium fraction, which is expected to be high in a pre-stellar core, and compare with their relative non sulphur bearing molecules. In Section 5, we use a detailed chemical modelling and confront the results with the observations. 

\section{Observations}

The observations for all transitions quoted in Table \ref{spectro} were performed at the IRAM-30m toward the dust peak emission of the L1544 pre-stellar core (${\rm \alpha_{2000} = 05^h04^m17.21^s, \delta_{2000} = 25\degr10\arcmin42.8\arcsec}$) in the framework of the ASAI Large Program, except for H$_2$S which was observed as a follow-up project. Observations at frequencies lower than 80 GHz have been performed in December 2015. All the details of these observations can be found in \citet{vastel2014} and \citet{quenard2017a} and line intensities are expressed in units of main-beam brightness temperature. \\
The ortho--H$_2$S (1$_{1,0}$ -- 1$_{0,1}$) at 168762.75 MHz has been observed with the IRAM-30m towards the dust peak emission on March 19 and 20, 2017, with the use of the spectral line Eight MIxer Receivers (EMIR) in band E150 combined with the narrow mode of the Fast Fourier Transform Spectrometers (FTS) allowing a spectral resolution of 50 kHz. We used the frequency switching observing mode with a frequency throw of 7.14 MHz, which allows a good removal of the ripples including standing waves between the secondary and the receivers \citep{fuente2016}. Pointing was checked every 1.5 hour on the nearby continuum sources 0430+352, 0439+360 and 0316+413 with errors always within 3$^{\prime\prime}$. The system temperature was stable, at $\sim$ 230 K (2 mm of precipitable water vapour) resulting in an average rms of 9.1 mK for a resolution of 50 kHz. The IRAM beam varies from 33.5$^{\prime\prime}$ at 75 GHz to 23.9$^{\prime\prime}$ at 105 GHz and 15$^{\prime\prime}$ at 168 GHz.\\

Twenty-one sulphur bearing species have been detected in total and are shown in Appendix A: CS, $^{13}$CS and C$^{34}$S (Fig. \ref{cs}), CCS and CC$^{34}$S (Fig. \ref{ccs}), C$_3$S (Fig. \ref{c3s}), H$_2$S (Fig. \ref{h2s}), H$_2$CS, H$_2$C$^{34}$S, HDCS and D$_2$CS (Fig. \ref{h2cs}), HSCN (Fig. \ref{hscn}), OCS (Fig. \ref{ocs}), SO, S$^{18}$O and $^{34}$SO (Fig. \ref{so}), SO$_2$ (Fig. \ref{so2}),  NS (Fig. \ref{ns}), NS$^+$ \citep{cernicharo2018}, HCS$^+$ and HC$^{34}$S$^+$(Fig. \ref{hcsp}). We present, in Table \ref{spectro}, the spectroscopic parameters of the transitions detected using the CDMS\footnote{http://www.astro.uni-koeln.de/} \citep{muller2005} for most species except CC$^{34}$S for which we used JPL\footnote{https://spec.jpl.nasa.gov/} \citep{pickett1998}. The line identification and analysis have been performed using the {\sc cassis}\footnote{http://cassis.irap.omp.eu} software \citep{vastel2015a}. The results from the line fitting take into account the statistical uncertainties accounting for the rms (estimated over a range of 15 km~s$^{-1}$ for a spectral resolution of 50 kHz). We also report a tentative detection (4 $\sigma$ level) for methyl mercaptan (CH$_3$SH) in Fig. \ref{ch3sh}.

\section{Determination of the column densities}

\begin{figure}
   \centering
   \includegraphics[width=1.06\hsize]{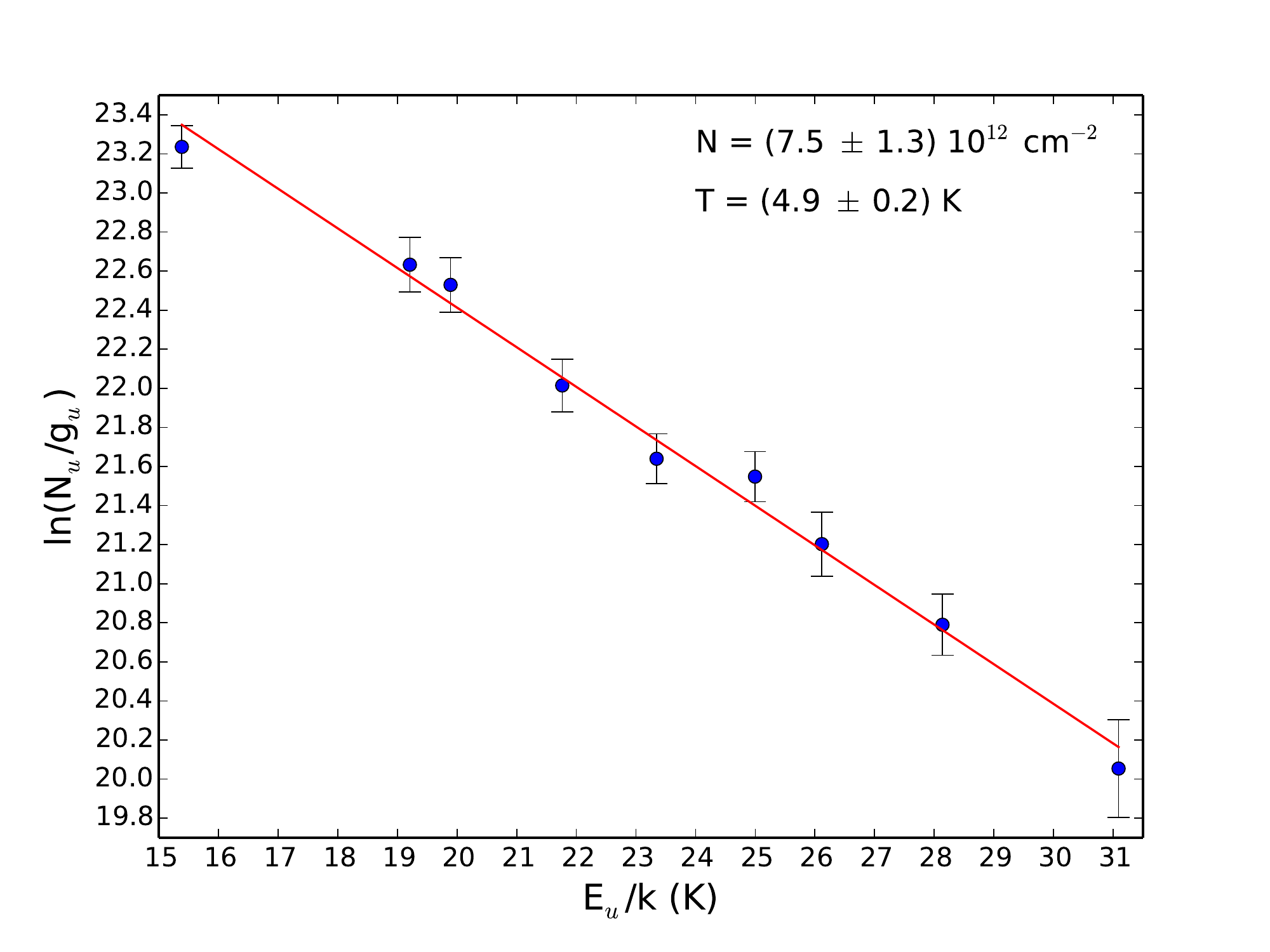}
   \caption{Rotational diagram analysis for the nine detected transitions of CCS. The CCS column density and rotational temperature are quoted in the upper right corner.}
   \label{RD}
 \end{figure}

In this section we determine the column densities (or some upper limits) for the detected species. Different methods are used, depending on the number of detected transitions for each species, and also depending on the availability of collisional coefficients. These methods take into account the uncertainties based on the line fitting and also the absolute calibration accuracy, around 10$\%$ or better depending on the band considered.
For species where multiple transitions have been detected, covering a wide range in energy, the rotational diagram method (see Fig. \ref{RD} in the case of CCS) is a useful tool to derive parameters such as the column density and excitation temperature. We performed this analysis for CCS, C$_3$S, H$_2$CS, H$_2$C$^{34}$S, HDCS, D$_2$CS, OCS and SO (see Table \ref{lte}). When multiple transitions are detected we can also use a MCMC (Markov Chain Monte Carlo) method implemented within  {\sc cassis}. The MCMC method is an iterative process that goes through all of the parameters with a random walk and heads into the solutions space and the final solution is given by a  $\chi^2$ minimization. All parameters such as column density, excitation temperature (or kinetic temperature and H$_2$ density in the case of a non-LTE\footnote{LTE: Local Thermodynamic Equilibrium} analysis), source size, linewidth, V$_{lsr}$, can be varied. The partition functions have been computed in {\sc cassis} for temperatures lower than 9.375 K which is the lowest temperature given by the CDMS database for some species. These are computed as:\\
\begin{equation}
Q(T) = \sum_{i} g_{i} \times exp(-E_{i}/kT) 
\end{equation}
where g$_i$ and E$_i$ are the statistical weight and energy, respectively, of the i level. Table \ref{lte} shows the results from the MCMC method for a LTE analysis. Note that the emission is compatible with a source size larger than the IRAM beam of $\sim$ 30$^{\prime\prime}$.

\begin{table*}
\caption{Temperature and column density of sulphur species in L1544, under the LTE condition, using MCMC and rotational diagram analysis.  \label{lte}}
\begin{tabular}{cccccccc}
\hline\hline
     & \multicolumn{4}{c}{MCMC}  & \multicolumn{2}{c}{Rotational diagram} \\
  \hline
    Species & $\rm T_{ex}$ & N                  & FWHM       & $\rm V_{lsr}$         & $\rm T_{rot}$ & N   \\
                  &   (K)         & (cm$^{-2}$)     & (km~s$^{-1}$)  & (km~s$^{-1}$) &     (K)       & (cm$^{-2})$  \\
  \hline
CCS & $5.6 \pm 0.2$ & $6.5(\pm 0.9)$ x $10^{12}$ & $0.40 \pm 0.01$&$7.18 \pm 0.01 $ & $4.9 \pm 0.2$&$7.5(\pm 1.3)$ x $10^{12}$ \\
C$_{3}$S & $6.2 \pm 0.5$ & $3.1(\pm 0.1)$ x $10^{12}$ & $0.42 \pm 0.01$ &$7.22 \pm 0.01 $ & $7.9 \pm 0.2$ & $8.8(\pm 0.7)$ x $10^{11}$ \\
H$_{2}$C$^{34}$S & $14.1 \pm 2.0$ & $3.5(\pm 0.7)$ x $10^{11}$ & $0.43 \pm 0.03$& $7.24 \pm 0.02$ & $14.6 \pm 0.8$ & $3.1(\pm 0.2)$ x $10^{11}$\\
D$_{2}$CS & $9.6 \pm 1.5$ & $1.1(\pm 0.7)$ x $10^{12}$ &$0.47 \pm 0.02$ &$7.22 \pm 0.01$ & $10.8 \pm 0.8$ & $6.5(\pm 0.6)$ x $10^{11}$ \\
OCS & $7.5 \pm 0.6$ & $6.3(\pm 1.6)$ x $10^{12}$ & $0.36 \pm 0.01$&$7.18 \pm 0.01$ & $8.8 \pm 0.2$ & $4.1(\pm 0.2)$ x $10^{12}$ \\
HDCS & $6.8 \pm 0.6$ & $1.6(\pm 0.8)$ x $10^{12}$ &$0.42 \pm 0.01$ & $7.23 \pm 0.01$& $7.5 \pm 0.5$ & $8.4(\pm 1.1)$ x $10^{11}$ \\
H$_{2}$CS & $12.3 \pm 0.7$ & $7.3(\pm 1.0)$ x $10^{12}$ &$0.41 \pm 0.01$ &$7.20 \pm 0.01$ & $13.1 \pm 0.9$ & $5.8(\pm 0.6)$ x $10^{12}$ \\
SO & $9.7 \pm 1.8$  & $5.2(\pm 0.8)$ x $10^{12}$ &$0.34 \pm 0.01$ &$7.22 \pm 0.01$  & $7.9 \pm 1.6$ & $7.2(\pm 3.2)$ x $10^{12}$ \\
\hline
\end{tabular} 
\end{table*}

 \begin{figure}
   \centering
   \includegraphics[width=1.06\hsize]{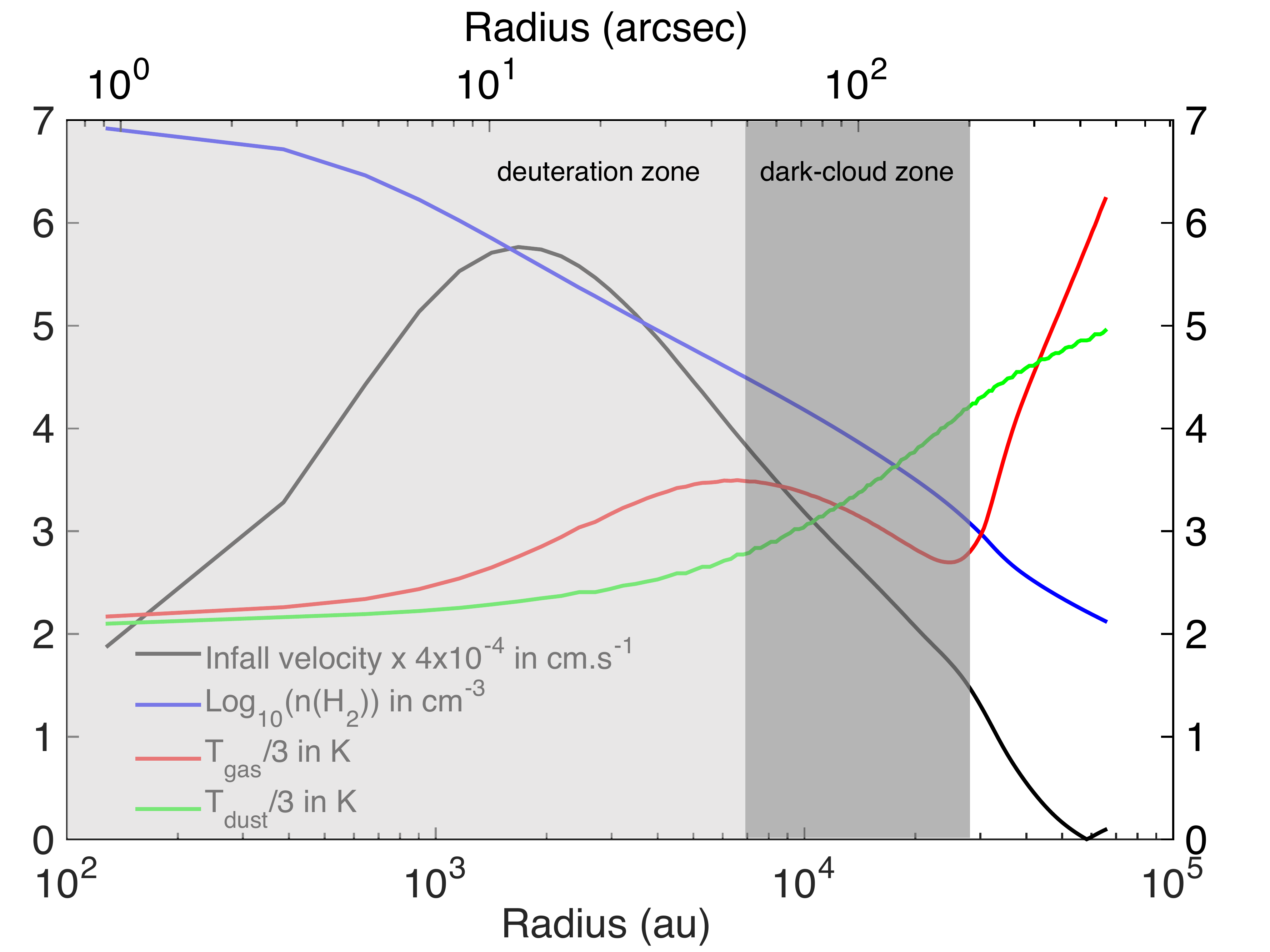}
   \caption{Gas and dust temperature, density, and velocity profiles of the L1544 pre-stellar core as a function of the radius in arcseconds and au from \citet{keto2014}.}
   \label{struct}
 \end{figure}

The computed column densities and temperatures shown in Table \ref{lte} show similar results, especially in the case of CCS for which we have nine transitions detected. Some differences occur when the number of transitions decreases. For example, there is a factor of two on the computation of the column density of HDCS (and a factor of three for C$_3$S) between both methods, reflecting the low number of transitions detected covering a small range in energy E$_{u}$ = 8.9, 17.7, 18.1 K (E$_{u}$ = 25.3, 29.1, 33.3, 33.7 K for C$_3$S). The rotational diagram analysis cannot be performed on species for which only a few transitions are detected. For these species, we fixed the excitation temperature at 10 K, which is the average kinetic temperature of the L1544 core in the IRAM beam (see Fig. \ref{struct}), or slightly varied the temperature to be compatible with the non detected transitions in our spectral survey, and constrained their column density by adjusting the spectra. The resulting column densities are given in Table \ref{lteat10K} as N$_{species}$. We present in Fig. \ref{ch3sh} the tentative detection for methyl mercaptan (CH$_3$SH) in our spectral survey, convincingly reinforced by a LTE modelling (in red) with a strong upper limit on the column density of  $2.5 \times 10^{11} cm^{-2}$ using a fixed temperature of 10 K, a full width at half maximum of 0.3 km/s, a velocity in the standard of rest of 7.1 km/s. Note that varying the excitation temperature from 8 to 12 K for these species does not significantly change the results.

\begin{table}
\caption{Derived column density (N) with a fixed excitation temperature of 10 K for species where one or two transitions have been detected or with a varied excitation temperature so that the LTE modelling is compatible with the non detected transitions in our spectral survey.  \label{lteat10K}}
\begin{tabular}{|c|c|c|c|}
  \hline\hline
    Species & $\rm T_{ex}$ & $\rm N_{species} $                     &   $\rm N_{main\,isotopologue}$\\
                  &   (K)              & (cm$^{-2}$)                             & (cm$^{-2}$)\\
	\hline
HCS$^{+}$& $10$ &  $(6.2-6.5) \times 10^{11}$                 &\\
HC$^{34}$S$^{+}$& $10$ &  $(4.8-5.2) \times 10^{10}$     & $(1.1-1.2) \times 10^{12}$\\
CS & $10$ &  $(4.0-4.4) \times 10^{12}$                             &\\
$^{13}$CS & $10$ &  $(2.6-3.0) \times 10^{11}$                 & $(1.8-2.0) \times 10^{13}$\\
C$^{34}$S  & 10 &  $(7.8-8.3) \times 10^{11}$                    & $(1.8-2.0) \times 10^{13}$\\
CC$^{34}$S   & 6--7  & $(4-6) \times 10^{11}$                    & $(0.9-1.4) \times 10^{13}$\\
HSCN & $10$ &  $(5.8-6.2) \times 10^{10}$                        & \\
NS & $10$ &  $(1.4-1.6) \times 10^{12}$                             &\\
$^{34}$SO & 5--6 & $ (1.3-1.6) \times 10^{12}$                  & $(3.0-3.6) \times 10^{13}$\\
S$^{18}$O & 6--8 & $ (3-3.2) \times 10^{11}$                     & $(1.7-1.8) \times 10^{14}$\\
CH$_3$SH & 10 & $\le 2.5 \times 10^{11}$                        & \\

\hline
\end{tabular} 
\end{table} 

For OCS, SO and SO$_2$ we carried out a non-LTE analysis using the LVG (Large Velocity Gradient) code by \citet{ceccarelli2003}, using the collision rates by \citet{green1978}, \citet{lique2007b} and  \citet{cernicharo2011} respectively. Table \ref{lvg} lists the results (H$_2$ density, kinetic temperature and column density) from the LVG analysis, taking into account the observed integrated fluxes and the corresponding rms (see Table \ref{spectro}). These species are likely emitted in the external layer where the density is lower and the temperature is higher than in the center where $\rm T_{gas} \sim T _{dust}$ $\sim$ 7 K and n(H$_2$) $\sim$ 10$^7$ cm$^{-3}$ (see Fig. \ref{struct}). Note also that the emission is compatible with a source size larger than the maximum IRAM beam of $\sim$ 30$^{\prime\prime}$. For NS$^+$, we used the column density derived from \citet{cernicharo2018}: 2.3 $\times$ 10$^{10}$ cm$^{-2}$.\\
The H$_2$S transition is the only transition among our sulphur bearing species that presents a double-peaked profile that cannot simply be analysed in LTE (see Fig. \ref{h2s}). We estimated a lower limit on the H$_2$S column density of 1.6 $\times$ 10$^{12}$ cm$^{-2}$, from a simple LTE modelling using an excitation temperature of 10 K and we use this limit in section 5 (see Fig. \ref{Ncol_density_1e2}) as a comparison for the chemical modelling. We also use, in Appendix B, the variation of H$_2$S abundance as a function of radius (from section 5) combined with a 3D radiative transfer treatment to try to reproduce the line profile, taking into account the density and temperature profiles as well as the velocity profile of the L1544 pre-stellar core.\\

\begin{table}
\caption{Results from the non-LTE analysis for the OCS, SO and SO$_2$ species.  \label{lvg}}
\begin{tabular}{|c|c|c|c|c|}
  \hline\hline
    Species & $\rm n_{H_2}$ & $\rm T_{K}$ & $\rm N$     \\
                  & (cm$^{-3}$)     &  (K)               & (cm$^{-2}$)  \\
	\hline
OCS& (7 $\pm$ 3.0) $\times$ $10^{3}$  & 13 $\pm$ 2 & 4 ($\pm$ 1) $\times 10^{12}$  \\
SO &  (2 $\pm$ 1) $\times$ $10^{4}$ & $\ge$ 12 & $\ge$ 8 $\times$ $10^{12}$  \\
SO$_2$ &  (2 $\pm$ 1) $\times$ $10^{4}$ & 12 $\pm$ 1 & (2.0--3.5) $\times$ $10^{12}$  \\
\hline
\end{tabular} 
\end{table}

Isotopologues may be used when transitions are optically thick. We decided to use the isotopologues of the more abundant species presented in Section 2 to refine some of the column densities determined earlier. For example, one transition for $^{12}$CS, $^{13}$CS and C$^{34}$S have been detected in our spectral survey. A simple LTE analysis gives a $^{12}$CS/$^{13}$CS ratio between 13.3 and 16.9, much below the value of 68 determined in the local interstellar medium \citep{milam2005,asplund2009,manfroid2009} for $^{12}$C/$^{13}$C. The $^{12}$CS (2--1) is likely optically thick and hinders the determination of the true column density for $^{12}$CS. Using $^{13}$CS and a $^{12}$C/$^{13}$C ratio of 68, we obtain a value of N($^{12}$CS)=(1.77--2.04) $\times$ 10$^{13}$ cm$^{-2}$. The same analysis can be also applied for $^{34}$S/$^{32}$S. The LTE analysis for C$^{34}$S gives a C$^{34}$S/C$^{32}$S ratio of $\sim$ 0.2, much higher than the value (0.044) in the vicinity of the Sun \citep{chin1996}. The CS column density derived from C$^{34}$S gives (1.77--1.89) $\times$ 10$^{13}$ cm$^{-2}$, similar to the value derived from $^{13}$CS. Previous observations of the CS (2--1) double peaked line profile, as well as a map over the whole core, have shown that CS is depleted in the central positions \citep{tafalla2002,hirota1998}. \citet{aikawa2003} used these observations to compute the column density from their best-fit model assuming a spherical core with radius of 15000 au. Using the collision coefficients for para-H$_2$ from \citet{green1978}, the resulting column density for the CS molecule is 4.6 $\times$ 10$^{13}$ cm$^{-2}$ (with an uncertainty factor of 2--3), a factor 10 higher than their LTE value, but compatible with our LTE computation using $^{13}$CS and $^{12}$CS/$^{13}$CS=68. The $^{34}$S/$^{32}$S ratio for both HCS$^+$ and CCS are compatible with the value found in the vicinity of the Sun. For the SO molecule, two isotopologues have been detected: $^{34}$SO and S$^{18}$O. The respective SO column densities based on the local values of the isotopic ratios are $\sim$ 3.2 $\times$ 10$^{13}$ and $\sim$ 1.8 $\times$ 10$^{14}$ cm$^{-2}$ respectively, compatible with the lower limit (8 $\times$ 10$^{12}$ cm$^{-2}$) found using a non-LTE formalism (see Table \ref{lvg}). \\
We present in the fourth column of Table \ref{lteat10K}, the column density (as $\rm N_{main\,isotopologue}$) of the main species based on the rarer isotopologue column density and using $^{12}$C/$^{13}$C, $^{32}$S/$^{34}$S and $^{16}$O/$^{18}$O ratios of 68, 23, and 557 \citep{wilson1999} respectively. CS and HCS$^+$ are the only species where optical depth affects the determination of the column densities and we will use their values, determined from the rare isotopologues ($^{13}$CS, C$^{34}$S and HC$^{34}$S$^+$), when comparing with the outcome of the chemical model described in Section 5.\\
Very recently, the thioformyl radical (HCS) and its metastable isomer HSC have been detected toward the molecular cloud L483 \citep{agundez2018}. These species have not been detected in our spectral survey, and we can estimate an upper limit on the column density for both species, using an excitation temperature of 10K: N(HCS) $\le$ 3 $\times$ 10$^{12}$ cm$^{-2}$ and N(HSC) $\le$ 6 $\times$ 10$^{10}$ cm$^{-2}$, a factor two to three lower than those derived towards L483.\\
Table \ref{final-N} summarizes the column densities that have been computed in this Section which will be used in Section 5.

\section{Deuterium Fraction of the sulphur bearing species}

An extreme molecular deuteration is a major characteristic of pre-stellar cores. Although the deuterium abundance is about 1.5 $\times$ 10$^{-5}$ relative to hydrogen (Linsky 2003), singly, doubly and even triply deuterated molecules have been detected with D/H ratios reaching 100$\%$ \citep[see][for a review]{ceccarelli2014}. Deuterium fractionation occurs in the cold and dense regions of the interstellar medium where CO is depleted from the gas phase, which leads to the preferential reactions between the H$_3$$^+$ ion with HD, feeding the deuterium of the latter ion, and distributing deuterium in the gas-phase and on the grain surfaces. In those regions where CO is highly depleted and H$_2$ is mostly in para form, the abundance of D$_2$H$^+$ should be similar to that of H$_2$D$^+$ \citep{roberts2003}. This was confirmed  with the detection of D$_2$H$^+$ toward the pre-stellar core 16293E \citep{vastel2004}. A high deuterium fractionation has already been detected in L1544 \citep[e.g.][]{caselli1999,crapsi2005,bizzocchi2014} in which H$_2$D$^+$ has been detected \citep{caselli2003}. The central 7000 au is called the {\it deuteration zone} (see Fig. \ref{struct}), where the freeze-out of abundant neutrals such as CO and O, the main destruction partners of the H$_3$$^+$ isotopologues, favour the formation of deuterated molecules. The outer ring (7000--30000 au), is called the {\it dark-cloud zone} \citep{caselli2012,ceccarelli2014}, where the carbon is mostly locked in CO, gas-phase chemistry is regulated by ion-molecule reactions and deuterium fractionation is reduced. \\
From Table \ref{lte} we can compute the following fractionation ratios: H$_2$C$^{34}$S/H$_2$CS = 0.048 $\pm$ 0.016, HDCS/H$_2$CS = 0.219 $\pm$ 0.140 and D$_2$CS/H$_2$CS = 0.151 $\pm$ 0.117. The first one is compatible with the $^{34}$S/$^{32}$S ratio of 0.044 in the vicinity of the Sun \citep{chin1996} which means that the H$_2$CS detected transitions are optically thin and give a good estimate of the total column density of H$_2$CS. The deuteration fractionation ratios for singly and doubly deuterated thioformaldehyde are both comparable and present an extremely high D enhancement. As a comparison, for the B1 cloud, the derived HDCS/H$_2$CS and D$_2$CS/H$_2$CS abundance ratios are 0.33 and 0.11 respectively \citep{marcelino2005}.\\ 
At the high densities and very low temperatures found in pre-stellar cores, D$_2$CO forms efficiently \citep{tielens1983} because of its lower zero energy level compared to that of H$_2$CO, through gas chemistry where deuterium is passed from the deuterated forms of H$_3^+$ and CH$_3^+$ \citep{roberts2003,roberts2007}. Considering the high densities in the central regions of L1544 where deuteration is the highest (see Fig. \ref{struct}), the LTE assumption should be correct for the determination of the column densities of the deuterated species, as found in the case of B1 \citep{marcelino2005}. The collision coefficients are unkown for the D$_2$CS and HDCS species, but assuming a typical range of 10$^{-11}$--10$^{-10}$ cm$^3$~s$^{-1}$ and using Einstein coefficients of $\sim$ 10$^{-5}$ s$^{-1}$(see Table \ref{spectro}), we can estimate critical densities between 10$^5$ and 10$^6$ cm$^{-3}$. These values correspond to the densities found at 4 $\times$ 10$^3$ and 10$^3$ au respectively from the L1544 center (see Fig. \ref{struct}) and are higher than densities found at larger distances. Therefore, it is reasonable to assume that LTE conditions are valid for the gas in the {\it deuteration zone}, whereas the emission from the outer gas is likely sub-thermal.\\
The formation of the deuterated forms of thioformaldehyde and formaldehyde should be similar \citep[CO and CS being strongly depleted:][]{tafalla2002} and their deuterated ratios comparable. A map of formaldehyde and its deuterated counterparts has recently been performed by Chac\`on-Tanarro et al. (submitted to A\&A) in L1544 and they measured the following deuterated fractions at the dust peak where CO is heavily depleted \citep{caselli1999}: D$_2$CO/H$_2$CO = 0.04 $\pm$ 0.03, HDCO/H$_2$CO = 0.03 $\pm$ 0.02. The derivation was done assuming optically thin emission and LTE, using a constant excitation temperature of 7 K (from the modelling of H$_2$CO). It is difficult to compare the deuterated fractions of both formaldehyde and thioformaldehyde because of the large error bars found for the latter. Overall, it seems that deuteration is somewhat more efficient for thioformaldehyde than for formaldehyde.
\\

\section{Evidence for sulphur depletion: a comparison between the observations and the chemical modelling}

\begin{figure*}
   \centering
   \includegraphics[width=0.95\hsize]{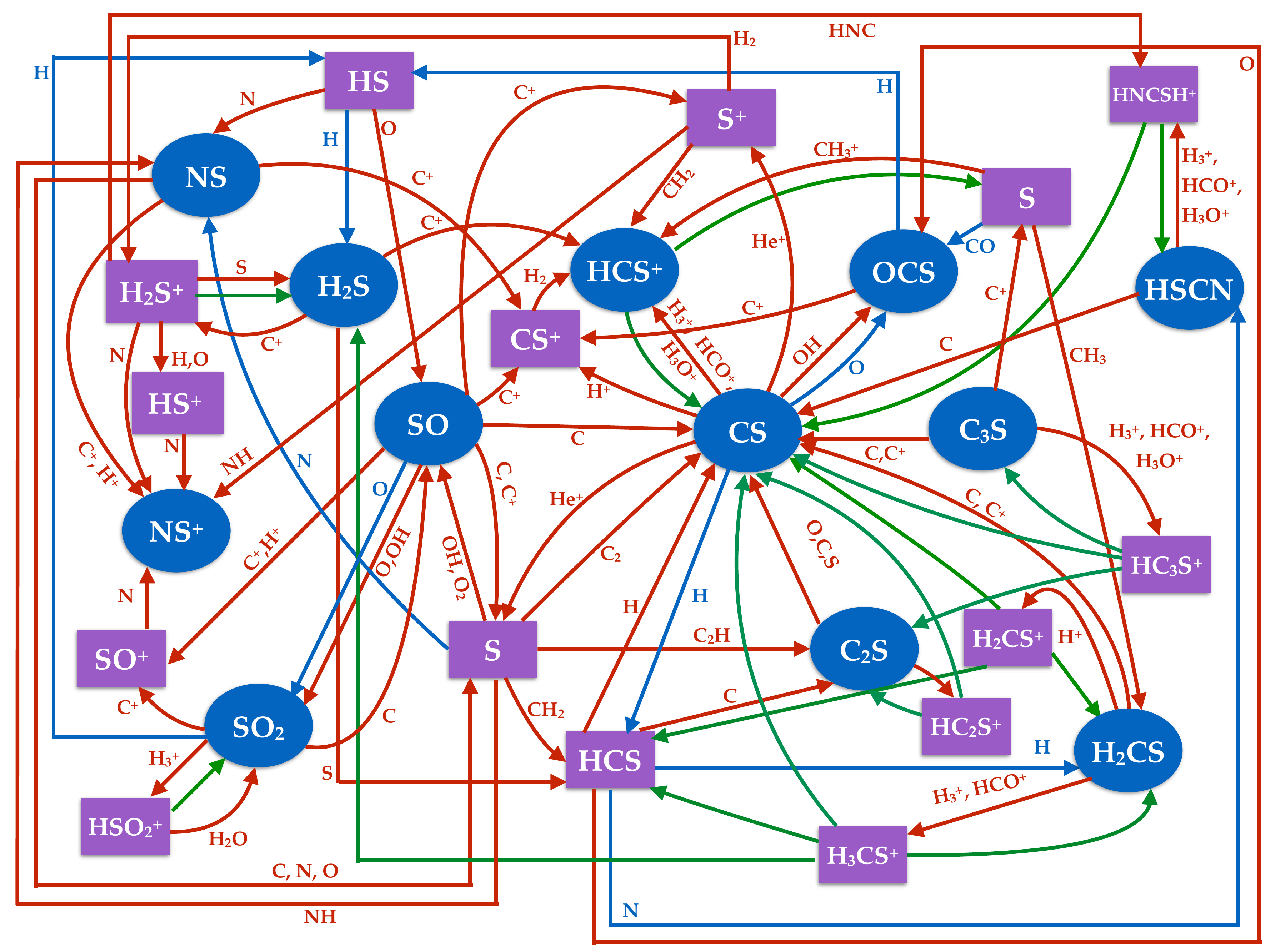}
   \caption{Simplified sulphur network. The blue ellipses being the detected species in L1544, the purple boxes being the intermediate undetected species, blue arrows are the surface reactions, green arrows the electronic recombination and red arrows the bimolecular reactions. }
   \label{network}
 \end{figure*}

We now confront the results from the radiative transfer modelling presented in section 3 with the output of a detailed chemical modelling. The sulphur chemical network has recently been enhanced by \citet{vidal2017}, using experimental and theoretical rates and branching ratios from the literature. Basically, they added 46 sulphur bearing species along with 478 reactions in the gas-phase, 305 reactions on the grain surface and 147 reactions in the grain bulk. They tested the effect for this updated network on the output of a gas-grain chemical model for dark clouds conditions, with different elemental sulphur abundances. Their results show that, depending on the age of the observed cloud, the sulphur reservoir could be either atomic sulphur in the gas phase or HS/H$_2$S in icy grain bulks. From the chemical modelling, they conclude that depletion of sulphur is not required to explain the observations of the TMC-1 dark cloud. This cloud is at an earlier stage than L1544, and presents a constant density ($\sim$ 10$^{4}$ cm$^{-3}$) and temperature ($\sim$ 10 K). \\
We used the same chemical network for our study combined with three-phases modellings, which allows to follow the evolution of chemical abundances for a given set of chemical and physical parameters. Gas-phase, grain surface and grain bulk chemistries are taken into account, along with exchanges between those phases: adsorption of gas-phase species onto the grain surfaces, thermal and non-thermal desorption of species from the grain surface into the gas-phase, and finally the exchange of species between the bulk and the surface of the grains. More details on the three-phase model can be found in \citet{ruaud2016} and \citet{vidal2017}. 
We present in Fig. \ref{network} the most critical reactions linking the sulphur bearing species that we detected (blue ellipses), but also the intermediate undetected species (purple boxes) and the reactions exchanges: blue arrows are for surface reactions, green arrows for electronic recombination and red arrows for bimolecular reactions. We used the {\sc nautilus}' outputs for the many models considered and extracted the main reactions leading to the production of the detected sulphur bearing species in L1544. These reactions are based on the KInetic Database for Astrochemistry (KIDA) (http://kida.obs.u-bordeaux1.fr/). 

We have used the gas-grain chemical code {\sc nautilus} in its three-phase model (gas phase, grain surface and mantle) to predict the abundances of all sulphur species in the cold core. We have already used {\sc nautilus} in previous studies of the chemistry involved in the cold core \citep{quenard2017a,vastel2018} and we follow here a similar method to stay consistent with these works. In the work led by \citet{vidal2017}, they considered a one-step chemical modelling with a constant density and temperature to model the physical conditions in TMC-1. Indeed, TMC-1 is a younger core than L1544 and the latter presents a density, temperature and velocity structure with evidence of gravitational contraction \citep{caselli2012,keto2014}. To take into account this structure, we therefore used a two-step model: the first phase represents the evolution of the chemistry in a diffuse or molecular cloud, with T = 20 K and several densities, ranging from 10$^2$ to 2 $\times$ 10$^4$ cm$^{-3}$, depending on the model considered \citep[see][for more details]{quenard2017a,vastel2018}. The initial abundances considered are those labelled as "EA1" in \citet{quenard2017a} and we only vary the sulphur atomic abundance (see below and Table \ref{init_dens}). We follow the chemistry in this phase during 10$^6$ years. We have checked that a variation of this age to 10$^5$, 5 $\times$ 10$^6$ or 10$^7$ years does not change the resulting column densities by more than a factor 1.5.
In order to compare the results from the chemical modelling (abundances with respect to H) with the observations (column density) of sulphur bearing molecules in L1544, we took into account the density profile across the core \citep[see][]{quenard2017a} to determine the column density from the chemical modelling instead of the abundance:\\

\begin{equation}
N(X) = 2 \times \sum_{i=2}^{n}(r_{i-1}-r_i) \times \frac{n(H)_{i-1}[X]_{i-1}+n(H)_i[X]_i}{2}
\end{equation}
where r is the radius from the center for every layer, n(H)$_i$ the gas density and [X]$_i$ the abundance at radius r$_i$. The different N(X) are then weigthed using a Gaussian function with a FWHM depending on the beam of the IRAM 30m telescope (between 15$^{\prime\prime}$ and 30$^{\prime\prime}$, depending on the frequency) to compare with the observations. This procedure was not adopted in the case of TMC-1 for which a constant density profile has been used. In the case of L1544, we cannot simply divide the observed column density by the total H$_2$ column density since the emission is not radially constant.  \\

\begin{table}
\caption{Column densities used for the comparison with the chemical modelling.  \label{final-N}}
\begin{tabular}{|c|c|c|}
\hline
Species & N                  & method \\
             &  (cm$^{-2}$)  &\\
\hline
CCS            & (0.9--1.4) $\times$ 10$^{13}$     &  CC$^{34}$S \\
C$_{3}$S    & 3.1 ($\pm$ 0.1) $\times$ 10$^{12}$     & MCMC \\
SO$_2$      & (2.0--3.5) $\times$ 10$^{12}$     & LVG \\
CS              & (1.8--2.0) $\times$ 10$^{13}$     & $^{13}$CS, C$^{34}$S \\
OCS           & 4 ($\pm$ 1) $\times$ 10$^{12}$  & LVG \\
H$_{2}$CS &  7.3 ($\pm$ 1.0) $\times$ 10$^{12}$     & MCMC\\
HSCN         & (5.8--6.2) $\times$ 10$^{10}$     & LTE  \\
NS              & (1.4--1.6) $\times$ 10$^{12}$     & LTE  \\
NS$^+$      & 2.3 $\times$ 10$^{10}$               & \citet{cernicharo2018}\\
HCS$^+$   & (1.1--1.2) $\times$ 10$^{12}$     & HC$^{34}$S$^+$  \\
SO             &  $\ge$ 8 $\times$ 10$^{12}$       & LVG\\
H$_2$S     & $\ge$ 1.6 $\times$ 10$^{12}$               & LTE \\
CH$_3$SH     & $\le$ 2.5 $\times$ 10$^{11}$               & LTE \\
\hline
\end{tabular} 
\end{table}

\begin{figure*}
   \centering
   \includegraphics[width=\hsize]{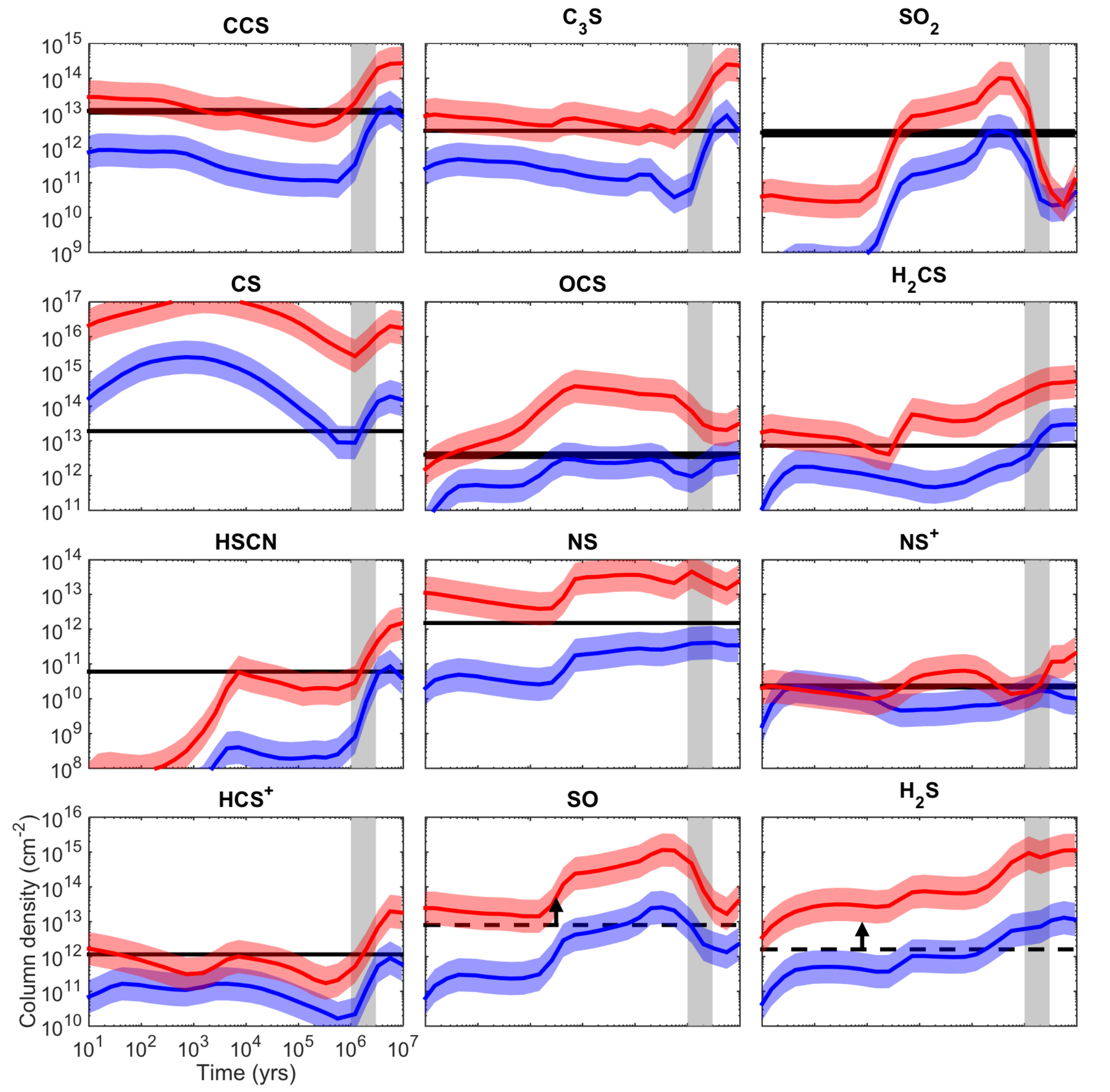}\\
   \caption{Variation of the modelled column density as a function of time and the comparison with the observed column density (black horizontal line). The blue lines correspond to model 1 (sulphur depletion: S/H=$8~10^{-8}$) and the red ones correspond to model 4 (sulphur non depletion: S/H=$1.5~10^{-5}$) as shown in Table \ref{init_dens}. The thickness of the black line corresponds to the error bar of the observed column densities. A variation by a factor of three of modelled column densities is shown in corresponding coloured areas. The dashed black horizontal line for SO and H$_2$S correspond to the lower limit on the computation of the total column density. The gray vertical area highlights an age between [1--3] $\times$ 10$^6$ years (see text).}
   \label{Ncol_density_1e2}
 \end{figure*}

We present in Fig. \ref{Ncol_density_1e2} the variation of the modelled column density as a function of time from \textsc{nautilus} and the comparison with the observed column density (black horizontal line). We use in Fig.  \ref{Ncol_density_1e2} (black horizontal lines) the observed column densities from Table \ref{final-N} that have been computed in Section 3 for the 13 sulphur bearing species. The width of the line represents the errors quoted in Table \ref{final-N}. The blue lines correspond to model 1 (sulphur depletion: S/H=8.0 $\times$ 10$^{-8}$) and the red ones correspond to model 4 (sulphur non depletion: S/H=1.5 $\times$ 10$^{-5}$) for an initial H density of 10$^2$ cm$^{-3}$. Table \ref{init_dens} shows the different modelling varying density in the first phase and the element gas phase abundances. From Fig. \ref{Ncol_density_1e2} we can clearly identify which model better reproduces the observations. A sulphur depletion seems necessary overall, with the exception of SO$_2$, although still within the error bars. Additional modelling are presented in Appendix C: Fig. \ref{Ncol_density_3e3} (initial H density of 3 $\times$ 10$^3$ cm$^{-3}$, see Table \ref{init_dens}) and \ref{Ncol_density_2e4} (initial H density of 2 $\times$ 10$^4$ cm$^{-3}$, see Table \ref{init_dens}).  

\begin{table}
	\centering
	\caption{Chemical modelling parameters used for the first phase model. \label{init_dens}}
	\begin{tabular}{llc}
	\hline\hline
	Model 	&	Density	&	Sulphur elemental\\
	number	&	(cm$^{-3}$)	&	abundance\\
	\hline
	1	&	$1\times10^2$	&	$8.0\times10^{-8}$\\
	2	&	$3\times10^3$	&	$8.0\times10^{-8}$\\
	3	&	$2\times10^4$	&	$8.0\times10^{-8}$\\
	4	&	$1\times10^2$	&	$1.5\times10^{-5}$\\
	5	&	$3\times10^3$	&	$1.5\times10^{-5}$\\
	6	&	$2\times10^4$	&	$1.5\times10^{-5}$\\
	\hline
	\end{tabular}
\end{table}

\begin{figure}
   \centering
   \includegraphics[width=0.9\hsize]{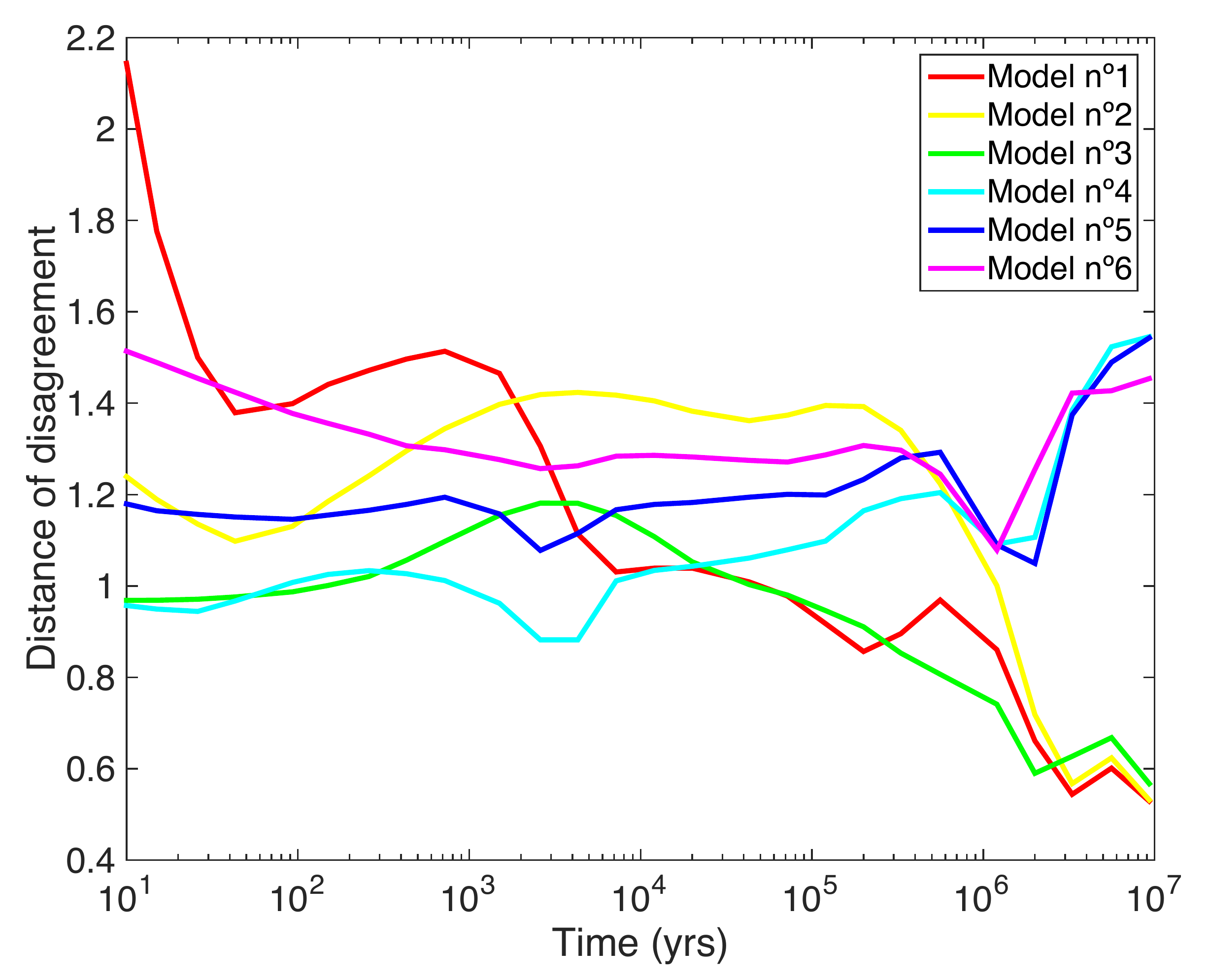}
   \caption{Distance of disagreement for models 1--3 (depletion) and 4--6 (non depletion). See Table \ref{init_dens}.}
   \label{disagreement}
 \end{figure}

\begin{figure*}
   \centering
   \includegraphics[width=0.7\hsize]{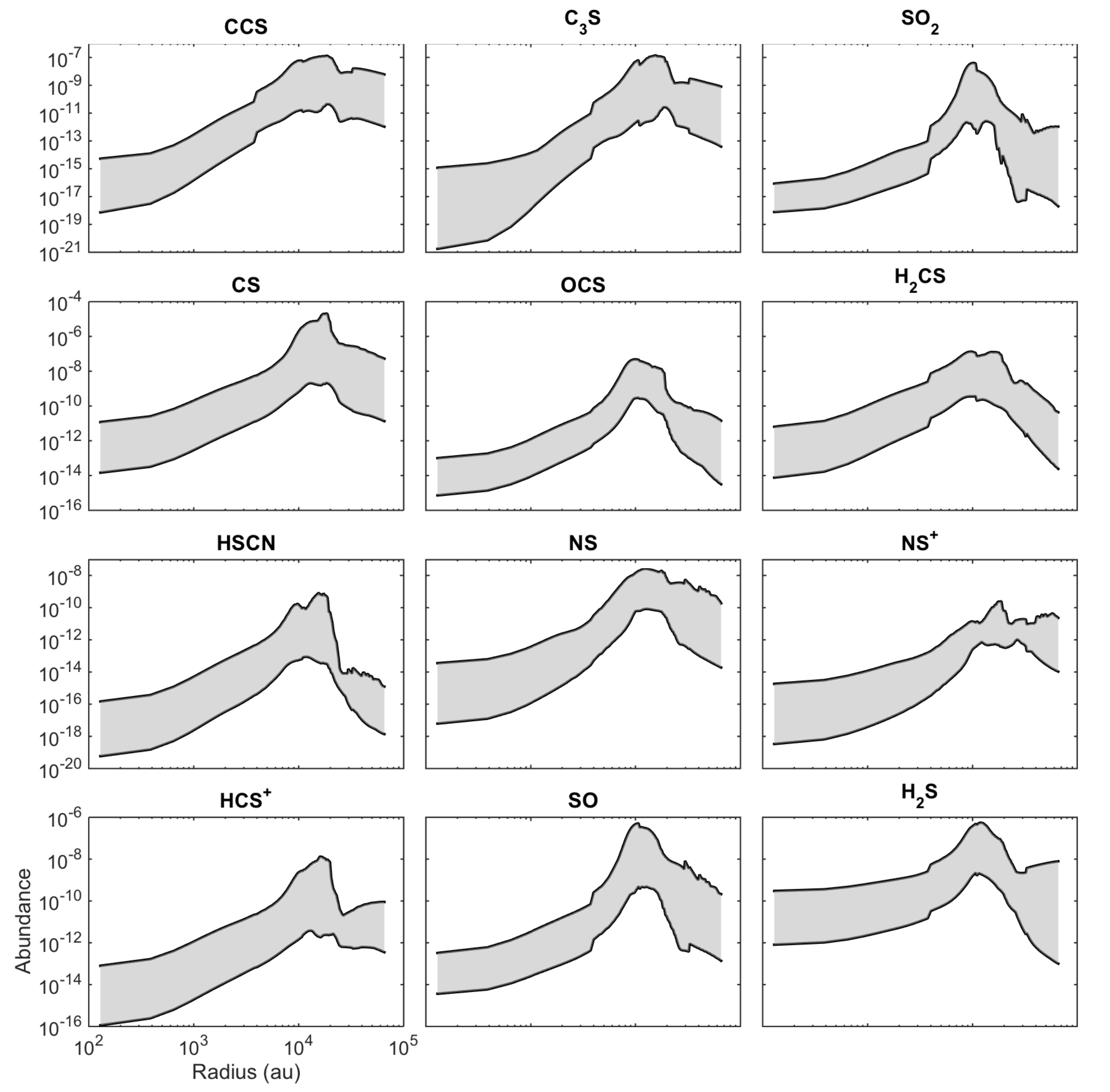}
   \caption{Radial distribution of the detected sulphur bearing species abundances in L1544 for an age between 10$^6$ and 3 $\times$ 10$^6$ years, for models 1--6 (see Table \ref{init_dens}).}
   \label{radial}
 \end{figure*}

Then, in order to find the "best-fit" model, we used the distance of disagreement computation \citep[see for example][]{wakelam2006}, applied on the column density, which is computed as follows:
\begin{equation}
D(t) = \frac{1}{n_{obs}} \sum_{i}|log(N(X))_{obs,i}-log(N(X))_{i}(t)|
\end{equation}
where N(X)$_{obs,i}$ is the observed column density, N(X)$_{i}$(t) is the modelled column density at a specific age and n$_{obs}$ is the total number of observed species considered in this computation (10 in the case of non deuterated sulphur bearing species detected in L1544). Note that we did not take into account species where a lower limit has been derived from the observations (SO and H$_2$S) and an upper limit has been derived (CH$_3$SH). Fig. \ref{disagreement} shows the distance of disagreement for all 6 models explained in Table \ref{init_dens}.\\
The minimum of the D(t) function is then obtained for the "best fit" age. Fig. \ref{disagreement} shows that 1) the "best-fit" age is 10$^6$-10$^7$ years for most models, compatible with previous estimates of the cloud age \citep{quenard2017a} and 2) the models where sulphur is depleted (models 1--3, respectively in red, yellow and green) are more favourable than models where sulphur is not depleted (models 4--6, respectively in light blue, dark blue and magenta). Moreover, the best solution for models 4 and 5 is found at very early age ($\sim$ 10$^3$ years), which is not compatible with the age of the object. We report the best-fit age in Fig. \ref{Ncol_density_1e2} as a gray vertical area which highlights an age between [1--3]~$\times 10^6$ years for a direct comparison between observations and modelling. The ten sulphur-bearing species are reproduced by the chemical model, within the observed error bars and considering a conservative factor of three to the modelled column densities. We are also in good agreement with the lower limits found for SO and H$_2$S.
We present in Fig. \ref{radial} the radial distribution for all the detected sulphur bearing species in the network from \citet{vidal2017}, for an age between 10$^6$ et 3 $\times$ 10$^6$ years, compatible with the results from the distance of disagreement. The abundances clearly peak at a radius between [1-2] $\times$ 10$^4$ au. Carbon, oxygen and sulphur depletion affect the cold and dense regions within L1544 \citep[e.g.][]{vasyunin2017}. Species like CO and CS disappear rapidly from the gas phase in the {\it deuteration zone} (see section 4), while species like N$_2$H$^+$ and NH$_3$ survive much longer at high densities. As a result, the pre-stellar core gradually develops a differentiated interior characterised by a centre rich in depletion-resistant species (such as the deuterium bearing species) surrounded by layers richer in depletion-sensitive molecules (such as the sulphur bearing species). This molecular differentiation as been identified in many starless cores \citep[e.g.][]{tafalla2006} and non-thermal desorption processes have been invoked \citep[e.g.][]{vastel2014,balucani2015,vastel2016,vasyunin2017}: FUV photo-desorption, cosmic-rays and chemical desorption.  As a consequence, we cannot simply assume a constant abundance as was assumed in dark clouds such as TMC-1 \citep{vidal2017} for the comparison between the observations and the chemical modelling. Note that in the current model, FUV photo-desorption plays a minor role as compared to the chemical desorption.\\

\begin{figure}
	\centering
	\includegraphics[width=0.8\hsize,clip=true,trim=0 0 0 1.29cm]{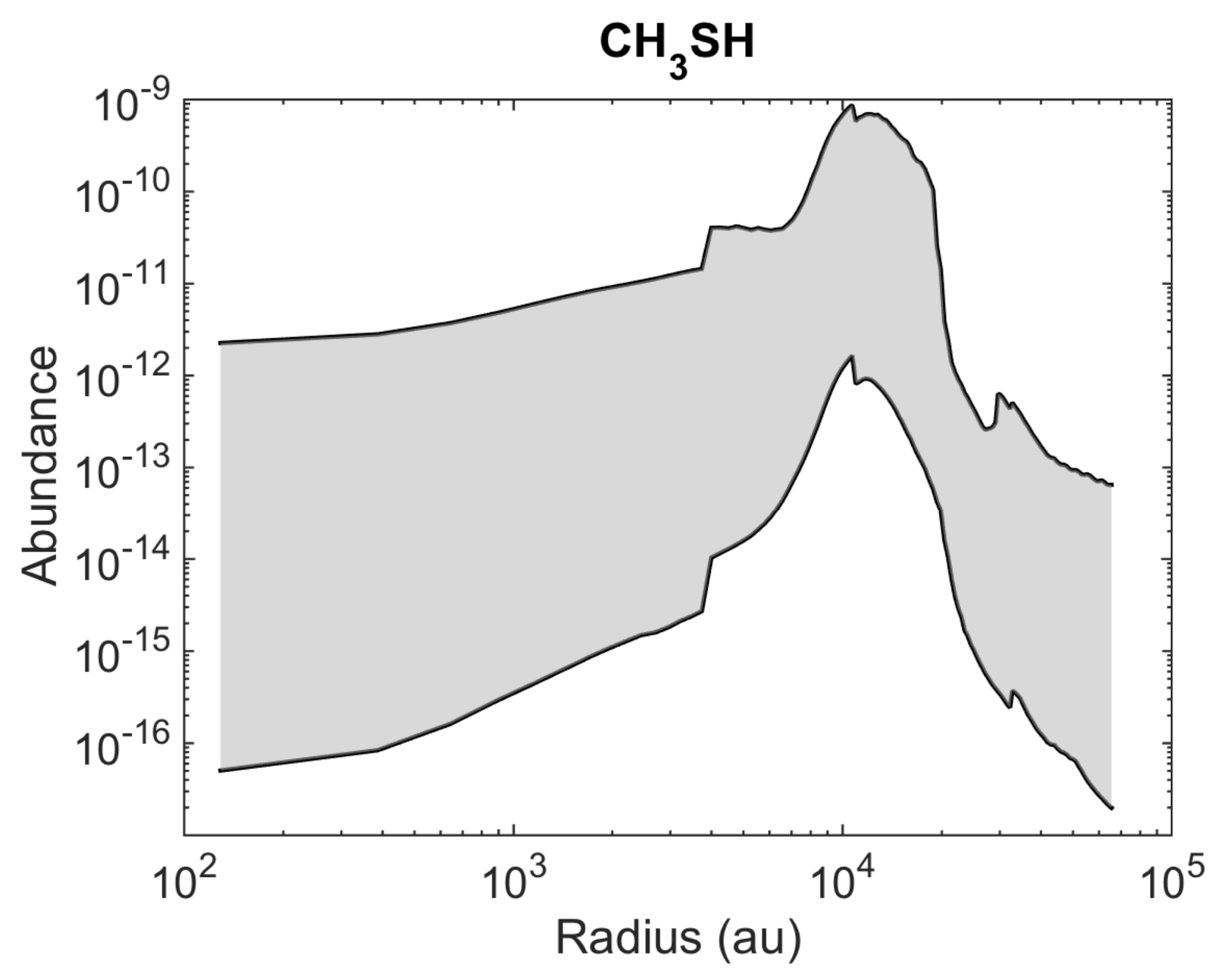}
	\includegraphics[width=0.8\hsize,clip=true,trim=0 0 0 0.9cm]{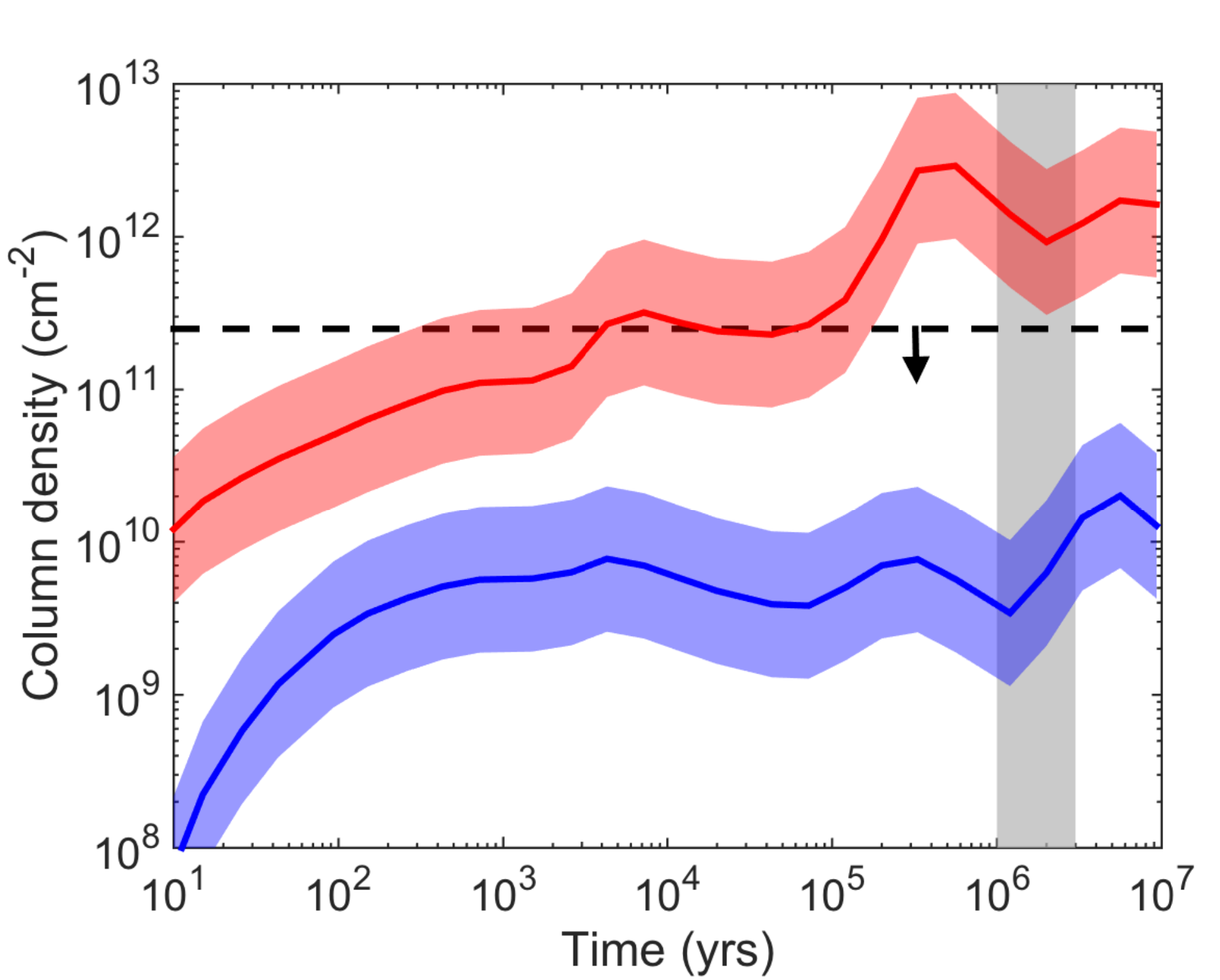}\\
	\includegraphics[width=0.8\hsize,clip=true,trim=0 0 0 0.9cm]{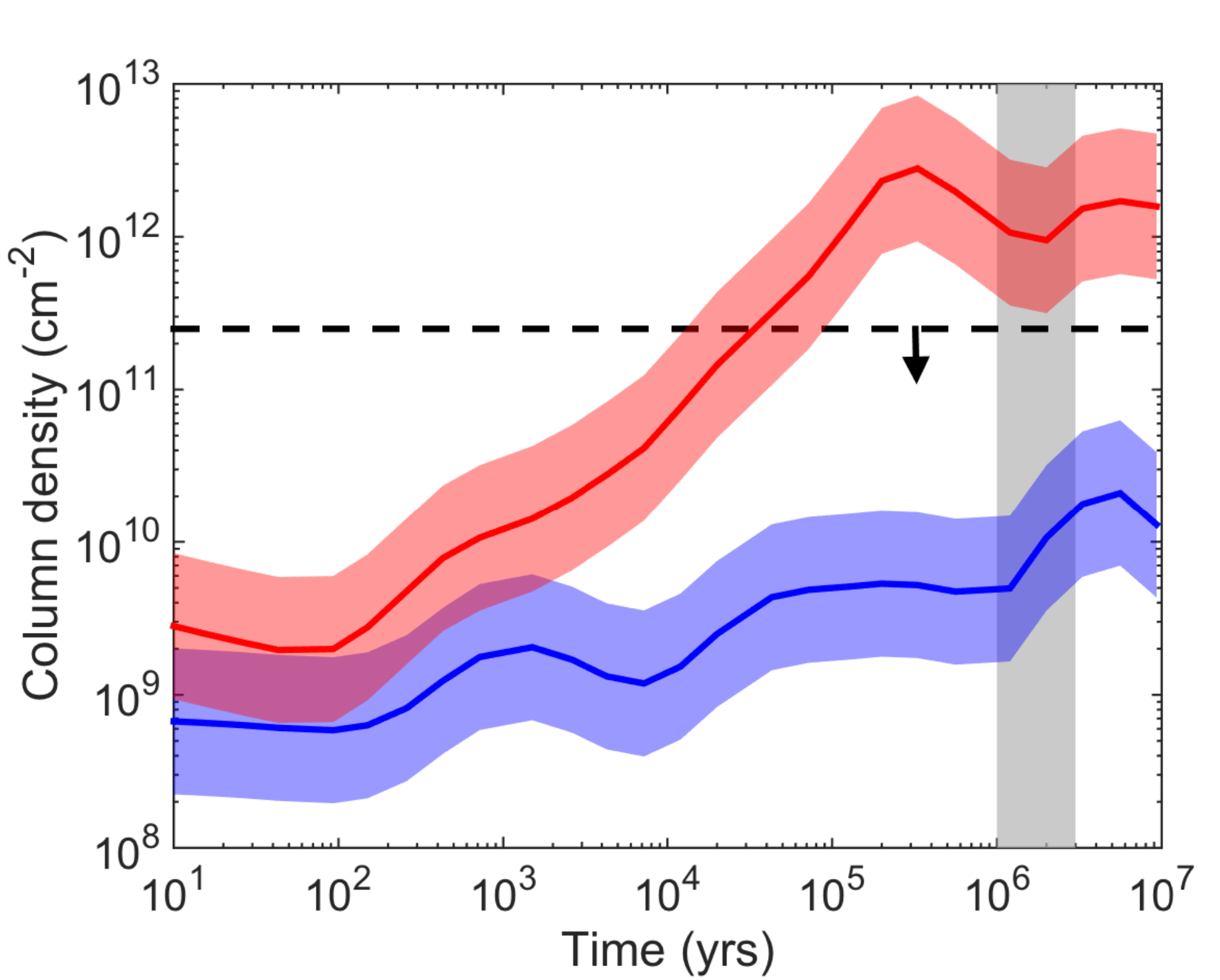}
	\includegraphics[width=0.8\hsize,clip=true,trim=0 0 0 0.9cm]{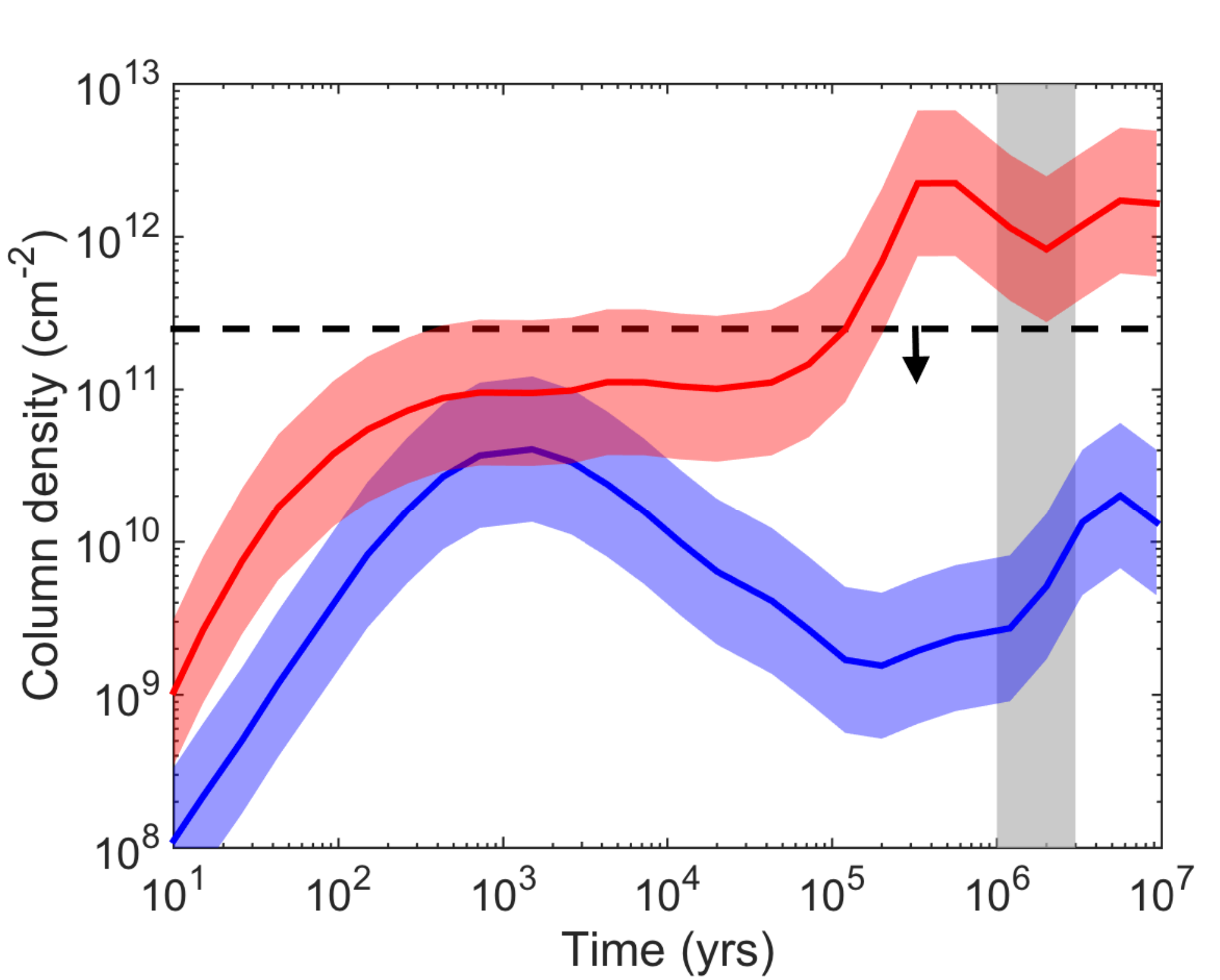}\\	
	\caption{Top panel: Radial distribution of CH$_3$SH for an age between 10$^6$ et 3 $\times$ 10$^6$ years, for models 1--6 (see Table \ref{init_dens}). Upper to lower panel: Variation of the modelled column density as a function of time and the comparison with the observed upper limit on the column density (dashed black horizontal line). The blue lines correspond to model 1, 2, 3 (sulphur depletion: S/H=$8~10^{-8}$) and the red ones correspond to model 4, 5, 6 (sulphur non depletion: S/H=$1.5~10^{-5}$) as shown in Table \ref{init_dens}. A variation by a factor of three of modelled column densities is shown in corresponding coloured areas. The gray vertical area highlights an age between [1--3] $\times$ 10$^6$ years (see text).}
	\label{ch3sh_model}
\end{figure}

Methyl mercaptan (CH$_3$SH) is tentatively detected in the L1544 pre-stellar core. It is likely emitted in the external layer of L1544 at $\sim$ 10$^4$ au (see Fig. \ref{ch3sh_model}) where the density drops to about 10$^4$ cm$^{-3}$ and the temperature increases to 10--12 K. We used the same chemical modelling as for the other sulphur-bearing species and present in Fig. \ref{ch3sh_model}  the results from the modelling for models 1--6 (see Table \ref{init_dens}) compared to the upper limit on the column density (dashed line). It is clear from this Figure that a non-depletion regime where S/H=$1.5~10^{-5}$ over produces this species (red line domain) and that a depletion regime where S/H=$8~10^{-8}$ (blue line domain) is more compatible with our observations. In the current network, CH$_3$SH is mainly formed in the gas phase (>80\%, depending on the model) through $\rm CH_3SH_2^+$ electron recombination ($\rm CH_3SH_2^+ + e^- \rightarrow CH_3SH + H$),  assuming that the $\rm CH_3^+~+~H_2S$ reaction leads to $\rm CH_3SH_2^+$.\\

To summarize, we used the ten detected species (CCS, C$_3$S, SO$_2$, CS, OCS, H$_2$CS, HSCN, NS, NS$^+$, HCS$^+$) and some of their isotopologues, as well as the lower limits on both SO and H$_2$S, and the upper limit on CH$_3$SH, to constrain the sulphur depletion in the L1544 pre-stellar core. The true degree of depletion of sulphur is difficult to constrain, but the results from our chemical modelling are consistent with a core where the initial sulphur elemental abundance is depleted with respect to the cosmic value (depletion $\sim$ 190). This is in contradiction with what was found by \citet{vidal2017} towards the starless core TMC-1 (CP). However, the method used by \citet{vidal2017} is very different from ours, since we are taking into account the density profile of the L1544 pre-stellar core. Their results might prove much different in case the density profile of TMC-1 (CP) is not flat. It is also difficult to compare our results with the study of the Barnard B1b globule which represents an advanced stage compared to L1544 and shows many structures in one beam. Higher spatial resolution is needed for Barnard B1b, to determine the possible influence of the B1b-S outflow on the sulphur chemistry.

\section{Conclusions}

We presented all the sulphur-bearing species that have been detected in a proto-typical pre-stellar core, L1544, using a spectral survey performed at the IRAM-30m. We computed the column densities for each species and compared them with the results from a chemical modelling taking into account the recent release from a sulphur chemical network. All species from this network are emitted in the external layer of the pre-stellar core, at about 10000 au, and the observations are best reproduced using an initial gas-phase sulphur abundance of 8 $\times$ 10$^{-8}$, 0.5$\%$ the cosmic value of 1.5~$\times$~10$^{-5}$. Sulphur is likely depleted in the cold and dense phases of the interstellar medium, although there is no strong evidence of sulphur in the observations of the ice mantles so far. 

\section*{Acknowledgements}
C.V. is grateful for the help of the IRAM staff at Granada during the data acquisitions, and also for their dedication to the telescope.




\bibliographystyle{mnras}
\bibliography{mnras_cvastel_V1.bbl} 


\newpage
\appendix
\section{Observations}

\begin{table*}
\tiny
\caption{Properties of the observed transitions. The spectroscopic parameters are from CDMS \citep{muller2005} except for NS$^+$ \citep{cernicharo2018}, and CC$^{34}$S (JPL). The rms has been computed over a range of 15 km~s$^{-1}$ with a spectral resolution of 50 kHz. \label{spectro} }
\begin{tabular}{|c|c|c|c|c|c|c|c|c|c|}
  \hline
    Species  & QN & Frequency (GHz) & $\rm E_{up} (K)$ & $\rm A_{ij}~(s^{-1})$ & rms (mK) & $\rm T_{mb}~(mK)$ & $\rm W~(mK~km~s^{-1})$ & $\rm FWHM~(km~s^{-1})$ & $\rm V_{LSR}$ \\
  \hline
CS                    & 20 -- 10 & $97.98095$ & $7.05$ & $1.68~10^{-5}$ & $3.3$ & $1226.5 \pm 0.1$ & $832.1 \pm 91.1$   & $0.64 \pm 0.07$ & $7.19 \pm 0.03$  \\
$^{13}$CS        & 20 -- 10 & $92.49431$ & $6.66$ & $1.41~10^{-5}$ & $2.3$ & $120.8 \pm 4.1$   & $56.3 \pm 3.2$       & $0.44 \pm 0.01$ & $7.28 \pm 0.01$  \\
C$^{34}$S        & 20 -- 10 & $96.41295$ & $6.94$ & $1.60~10^{-5}$ & $3.2$ & $391.4 \pm 4.2$   &  $156.6  \pm 21.3$ & $0.40 \pm 0.05$ & $7.16 \pm 0.01$  \\
CCS 		& 6$_7$ -- 5$_6$ & $81.50517$ & $15.39$ & 2.43~$ 10^{-5}$ & $6.8$ & $725.9 \pm 12.7$ & $346.3 \pm 13.8$ & $0.45 \pm 0.01$ & $7.07 \pm 0.01$ \\  
			& 6$_5$ -- 5$_4$ & $72.32379$ & $19.21$ & 1.60~$ 10^{-5}$ & $8.5$ & $261.7 \pm 12.7$ &$116.5 \pm 11.2$ & $0.42 \pm 0.02$ & $7.15 \pm 0.01$  \\
			& 7$_8$ -- 6$_7$ & $93.87011$ & $19.89$ & $3.74$~$ 10^{-5}$ & $3.2$ & $472.3 \pm 21.9$ & $225.3 \pm 21.7$ & $0.45 \pm 0.02$ & $7.23 \pm 0.01$ \\
			& 6$_6$ -- 5$_5$ & $77.73171$ & $21.76$ & 2.03~$ 10^{-5}$ & $5.2$ & $179.2 \pm 8.4$ & $85.5 \pm 7.8$ & $0.45 \pm 0.02$ & $7.15 \pm 0.01$ \\
			& 7$_6$ -- 6$_5$ & $86.18139$ & $23.35$ & 2.78~$ 10^{-5}$ & $3.4$ & $137.2 \pm 4.9$ & $62.5 \pm 5.1$ & $0.43 \pm 0.02$ & $7.13 \pm 0.01$ \\
			& 8$_9$ -- 7$_8$ & $106.34773$ & $25.00$ & $5.48$~$ 10^{-5}$ & $4.2$ & $219.6 \pm 7.6$ & $109.4 \pm 8.4$ & $0.47 \pm 0.02$ & $7.16 \pm 0.01$ \\
			& 7$_7$ -- 6$_6$ & $90.68638$ & $26.12$ & $3.29$~$ 10^{-5}$ & $3.5$ & $113.7 \pm 6.7$ & $54.2 \pm 6.8$ & $0.45 \pm 0.03$ & $7.18 \pm 0.01$ \\
			& 8$_7$ -- 7$_6$ & $99.86652$ & $28.14$ & $4.40$~$ 10^{-5}$ & $2.6$ & $75.5 \pm 4.5$ & $40.8 \pm 4.8$ & $0.51 \pm 0.03$ & $7.22 \pm 0.02$ \\
			& 8$_8$ -- 7$_7$ & $103.64076$ & $31.09$ & $4.90$~$ 10^{-5}$ & $5.2$  & $53.5 \pm 6.2$ & $25.0 \pm 5.7$ & $0.44 \pm 0.05$ & $7.15 \pm 0.02$ \\
CC$^{34}$S     & 6$_7$ -- 5$_6$  & $79.827457$ & $15.08$ & $2.31$~$ 10^{-5}$ & $4.8$  & $38.3 \pm 4.4$ & $18.7 \pm 4.6$ & $0.46 \pm 0.06$ & $7.30 \pm 0.03$ \\
                        & 7$_8$ -- 6$_7$  & $91.9135291$ & $19.49$ & $3.56$~$ 10^{-5}$ & $2.0$  & $21.5 \pm 1.9$ & $9.1 \pm 1.7$ & $0.40 \pm 0.04$ & $7.24 \pm 0.02$ \\
C$_3$S            & 13 -- 12 & 75.14793 & 25.25 & 3.26~10$^{-5}$ & $3.0$ & $108.7 \pm 4.7$ & $49.5 \pm 4.4$ & $0.43 \pm 0.02$ & $7.26 \pm 0.01$  \\
			& 14 -- 13 & 80.92818 & 29.13 & 4.09~10$^{-5}$ & $5.4$ & $72.5 \pm 7.8$ & $32.3 \pm 7.3$ & $0.42 \pm 0.05$ & $7.17 \pm 0.02$ \\
			& 15 -- 14 & 86.70838 & 33.29 & 5.04~10$^{-5}$ & $2.5$ & $76.5 \pm 3.1$ & $48.7 \pm 4.4$ & $0.60 \pm 0.03$ & $7.43 \pm 0.01$ \\
			& 16 -- 15 & 92.48849 & 37.73 & 6.13~10$^{-5}$ & $2.5$ & $23.7 \pm 2.0$ & $10.8 \pm 1.9$ & $0.43 \pm 0.04$ & $7.22 \pm 0.02$ \\ 
H$_2$S  & 1$_{1,0}$ -- 1$_{0,1}$ & 168.76276 & 27.88 &  $2.68$~$10^{-5}$ & 9.1 &  & $179.8 \pm 3.3$ & & \\
H$_2$CS  & 3$_{1,3}$ -- 2$_{1,2}$ & $101.4778$ & $22.91$ & $1.26$~$ 10^{-5}$ & $4.0$ & $558.4 \pm 11.7$ &$260.4 \pm 11.4$ & $0.44 \pm 0.01$ & $7.19 \pm 0.01$  \\
			& 3$_{0,3}$ -- 2$_{0,2}$ & $103.0405$ & $9.89$ & $1.48$~$ 10^{-5}$ & $5.6$ & $536.9 \pm 16.5$ & $256.1 \pm 19.3$ & $0.45 \pm 0.02$ & $7.20 \pm 0.01$ \\
			& 3$_{1,2}$ -- 2$_{1,1}$ & $104.6170$ & $23.21$ & $1.38$~$ 10^{-5}$ & $4.1$ & $514.3 \pm 12.0$ & $239.9 \pm 11.0$ & $0.44 \pm 0.01$ & $7.22 \pm 0.01$ \\
H$_2$C$^{34}$S  & 3$_{1,3}$ -- 2$_{1,2}$ & 99.77412 & 22.76 & 1.20~10$^{-5}$ & 3.1 & 24.8 $\pm$ 1.6 & $12.6 \pm 1.6$ & 0.48 $\pm$ 0.03 & 7.31 $\pm$ 0.01  \\
			& 3$_{0,3}$- - 2$_{0,2}$ & 101.2843 & 9.72 & 1.41~10$^{-5}$ & 3.2 & 26.5 $\pm$ 3.0 & $12.1 \pm 3.3$ & 0.43 $\pm$ 0.07 & 7.20 $\pm$ 0.03 \\
			& 3$_{1,2}$ -- 2$_{1,1}$ & 102.8074 & 23.05 & 1.31~10$^{-5}$ & 4.9 & 26.1 $\pm$ 3.1 & $11.6 \pm 2.8$ & 0.42 $\pm$ 0.05 & 7.20 $\pm$ 0.02 \\
HDCS  & 3$_{1,3}$ -- 2$_{1,2}$ & $91.17107$ & $17.73$ & $9.12$~$ 10^{-6}$ & $3.0$ & $47.2 \pm 2.3$ &$22.5 \pm 2.1$ & $0.45 \pm 0.02$ & $7.16 \pm 0.01$  \\
			& 3$_{0,3}$ -- 2$_{0,2}$ & $92.98160$ & $8.93$ & $1.09$~$ 10^{-5}$ & $2.7$ & $157.8 \pm 7.7$ & $76.9 \pm 7.1$ & $0.46 \pm 0.02$ & $7.23 \pm 0.01$ \\
			& 3$_{1,2}$ -- 2$_{1,1}$ & $94.82849$ & $18.09$ & $1.03$~$ 10^{-5}$ & $3.0$ & $38.5 \pm 4.6$ & $18.4 \pm 3.0$ & $0.45 \pm 0.02$ & $7.16 \pm 0.01$ \\
D$_2$CS  & $313 -- 212$ & $83.07776$ & $14.32$ & $7.00$~$ 10^{-6}$ & $3.9$ & $19.2 \pm 2.6$ &$11.6 \pm 3.4$ & $0.57 \pm 0.09$ & $7.30 \pm 0.04$  \\
			& $303 -- 202$ & $85.15392$ & $8.18$ & $8.48$~$ 10^{-6}$ & $4.0$ & $85.2 \pm 4.7$ & $44.3 \pm 4.3$ & $0.49 \pm 0.02$ & $7.20 \pm 0.01$ \\
			& $312 -- 211$ & $87.3027$ & $14.72$ & $8.12$~$ 10^{-6}$ & $6.3$ & $19.3 \pm 2.0$ & $8.6 \pm 3.6$ & $0.42 \pm 0.13$ & $7.17 \pm 0.05$ \\
			& $414 -- 313$ & $110.7561$ & $19.63$ & $1.81$~$ 10^{-5}$ & $5.0$ & $27.1 \pm 3.1$ & $13.2 \pm 2.9$ & $0.46 \pm 0.05$ & $7.29 \pm 0.02$ \\
HSCN  & 8$_{0,8}$ -- 7$_{0,7}$ & $91.75064$ & $19.82$ & $4.69$~$ 10^{-5}$ & $2.1$ & $14.3 \pm 1.5$ &$5.9 \pm 1.0$ & $0.39 \pm 0.03$ & $7.21 \pm 0.01$  \\
OCS  & $6 -- 5$ & $72.97678$ & $12.26$ & $1.07$~$ 10^{-6}$ & $3.5$ & $106.3 \pm 6.8$ &$40.6 \pm 3.7$ & $0.36 \pm 0.01$ & $7.18 \pm 0.01$  \\
			& $7 -- 6$ & $85.13910$ & $16.34$ & $1.71$~$ 10^{-6}$ & $4.1$ & $87.4 \pm 7.2$ & $35.2 \pm 3.8$ & $0.38 \pm 0.01$ & $7.19 \pm 0.01$ \\
			& $8 -- 7$ & $97.30121$ & $21.01$ & $2.58$~$ 10^{-6}$ & $4.7$ & $70.5 \pm 5.4$ & $27.7 \pm 2.9$ & $0.37 \pm 0.01$ & $7.19 \pm 0.01$ \\
			& $9 -- 8$ & $109.4631$ & $26.27$ & $3.70$~$ 10^{-6}$ & $4.6$ & $49.4 \pm 6.4$ & $17.8 \pm 4.4$ & $0.34 \pm 0.04$ & $7.15 \pm 0.02$ \\
SO                & $22 -- 11$ & $86.09395$ & $19.31$ & $5.25$~$ 10^{-6}$ & $3.2$ & $223.7 \pm 5.9$ &$100.0 \pm 5.0$ & $0.42 \pm 0.01$ & $7.162 \pm 0.002$  \\
	             & $23 -- 12$ & $99.29987$ & $9.23$   & $1.13$~$ 10^{-5}$ & $2.8$ & $1422.5 \pm 40.6$ & $678.5 \pm 34.4$ & $0.45 \pm 0.01$ & $7.10 \pm 0.01$ \\
		     & $32 -- 21$ & $109.2522$ & $21.05$ & $1.08~10^{-5}$ & $3.9$ & $176.1 \pm 5.2$ & $72.8 \pm 4.0$ & $0.39 \pm 0.01$ & $7.31 \pm 0.01$ \\
S$^{18}$O    & $23 -- 12$  & $93.26727$ & $8.72$   & $9.34~10^{-6}$ & $2.6$ & $21.3 \pm 2.1$  & $55.9 \pm 2.5$  & $0.36 \pm 0.02$ & $6.85 \pm 0.01$\\ 
$^{34}$SO  & $23 -- 12$ & $97.71539$  & $9.09$  & $1.09~10^{-5}$  & $4.4$ & $85.2 \pm 3.9$   & $223.3 \pm 4.1$ &$0.36 \pm 0.01$ & $6.93 \pm 0.01$\\
SO$_2$  & 3$_{1,3}$ -- 2$_{0,2}$ & 104.0294 & $7.74$ & $1.01$~$ 10^{-5}$ & $6.9$ & $153.7 \pm 8.7$ &$60.3 \pm 6.7$ & $0.37 \pm 0.02$ & $7.20 \pm 0.01$  \\
			& 6$_{0,6}$ -- 5$_{1,5}$ & 72.75824 & 19.16 & 2.77~10$^{-6}$ & 4.5 & 110.23 $\pm$ 0.01 & $34.6 \pm 3.9$ & 0.34 $\pm$ 0.02 & 7.20 $\pm$ 0.01 \\
NS  & 2$\pi_{1/2}$ 5/2$_{1,7/2}$ -- 3/2$_{-1,5/2}$ & $115.15393$ & $8.84$ & $2.33$~$ 10^{-5}$ & $15.1$ & $158.7 \pm 8.9$ &$74.0 \pm 7.5$ & $0.44 \pm 0.02$ & $7.35 \pm 0.01$  \\
		& 2$\pi_{1/2}$ 5/2$_{1,5/2}$ -- 3/2$_{-1,3/2}$ & $115.15681$ & $8.84$ & $1.96$~$ 10^{-5}$ & $15.0$ & $108.5 \pm 18.2$ & $41.4 \pm 17.3$ & $0.36 \pm 0.09$ & $7.38 \pm 0.04$ \\
		& 2$\pi_{1/2}$ 5/2$_{1,3/2}$ -- 3/2$_{-1,1/2}$ & $115.16298$ & $8.84$ & $1.75$~$ 10^{-5}$ & $14.6$ & $61.1 \pm 9.0$ & $34.3 \pm 9.6$ & $0.53 \pm 0.07$ & $7.54 \pm 0.03$ \\
		& 2$\pi_{1/2}$ 5/2$_{-1,7/2}$ -- 3/2$_{1,5/2}$ & $115.55625$ & $8.90$ & $2.35$~$ 10^{-5}$ & $17.8$ & $129.7 \pm 7.4$ & $49.5 \pm 5.6$ & $0.36 \pm 0.02$ & $6.95 \pm 0.01$ \\
NS$^+$   & 2 -- 1   & 100.19855   & 7.22  & $2.21~10^{-5}$ & 3.4  & 13.7 $\pm$ 1.3  & 8.6 $\pm$ 1.8  & 0.59 $\pm$ 0.07 & 6.95 $\pm$ 0.03\\
HCS$^+$  & 2 -- 1 & $85.34789$ & $6.14$ & $1.11$~$ 10^{-5}$ & $3.8$ & $246.8 \pm 6.1$ &$112.5 \pm 5.4$ & $0.43 \pm 0.01$ & $7.33 \pm 0.01$  \\
HC$^{34}$S$^+$  & 2 -- 1 & $83.96563$ & $6.04$ & $1.06$~$ 10^{-5}$ & $4.1$ & $22.4 \pm 4.0$ &$8.6 \pm 3.2$ & $0.36 \pm 0.07$ & $7.35 \pm 0.03$  \\

\hline
\end{tabular} 
\end{table*}

  \begin{figure}
   \centering
   \includegraphics[width=7cm]{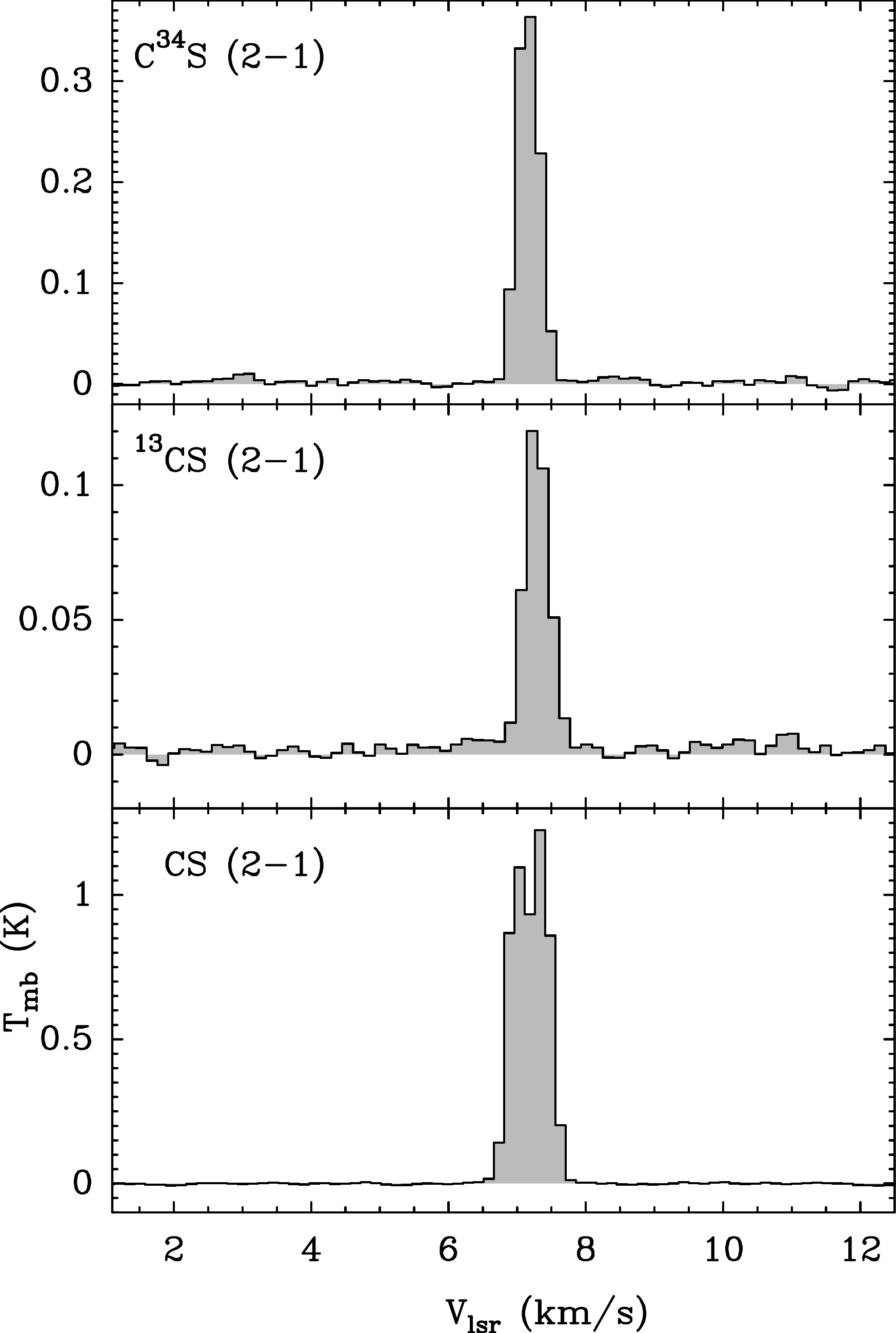}
   \caption{$^{12}$CS, $^{13}$CS and C$^{34}$S 2--1 detected line (in $\rm T_{mb}$).}
   \label{cs}
 \end{figure}
 
 \begin{figure}
   \centering
   \includegraphics[width=7cm]{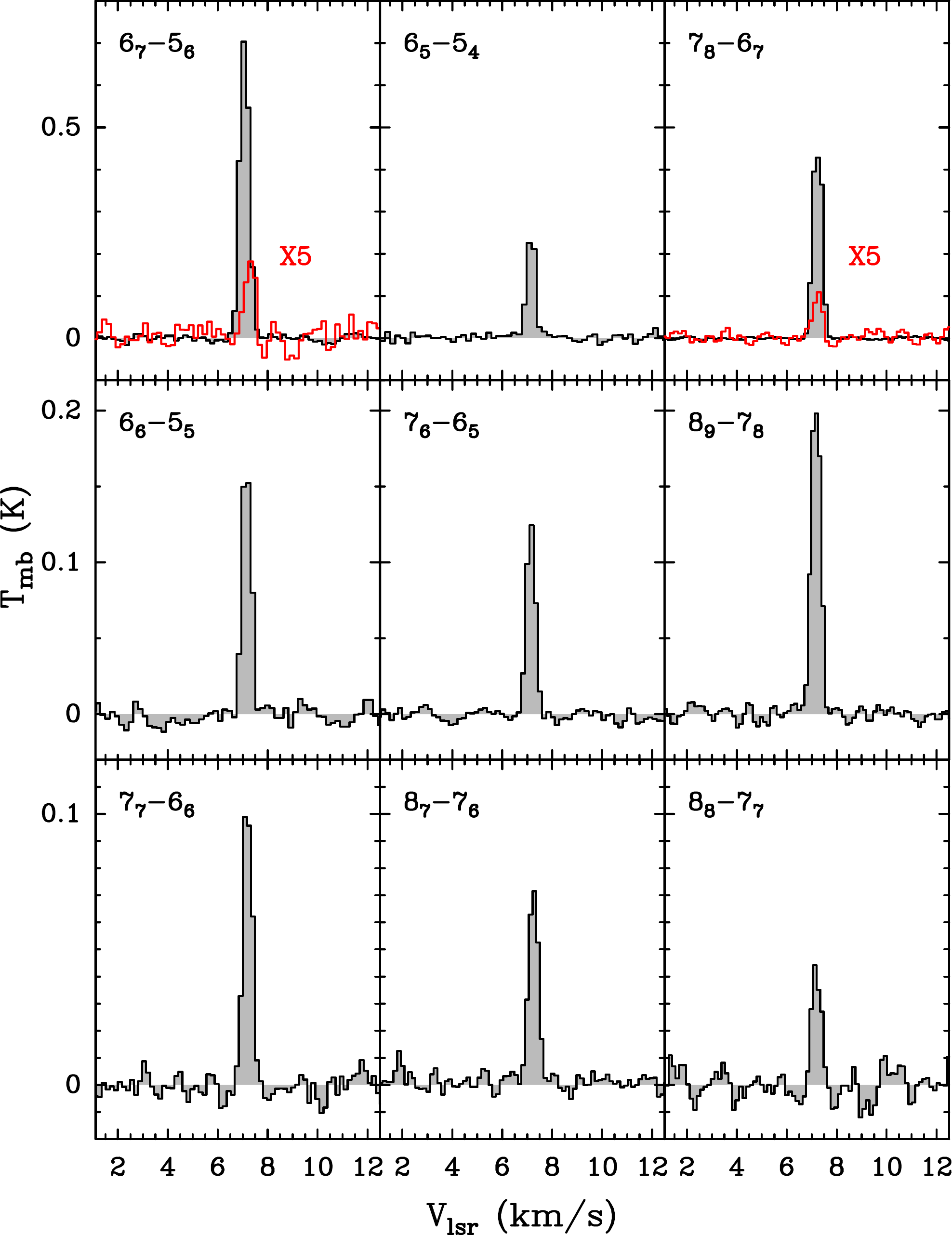}
   \caption{CCS (in black) and CC$^{34}$S (in red) detected lines (in $\rm T_{mb}$).}
   \label{ccs}
 \end{figure}

 \begin{figure}
   \centering
   \includegraphics[width=7cm]{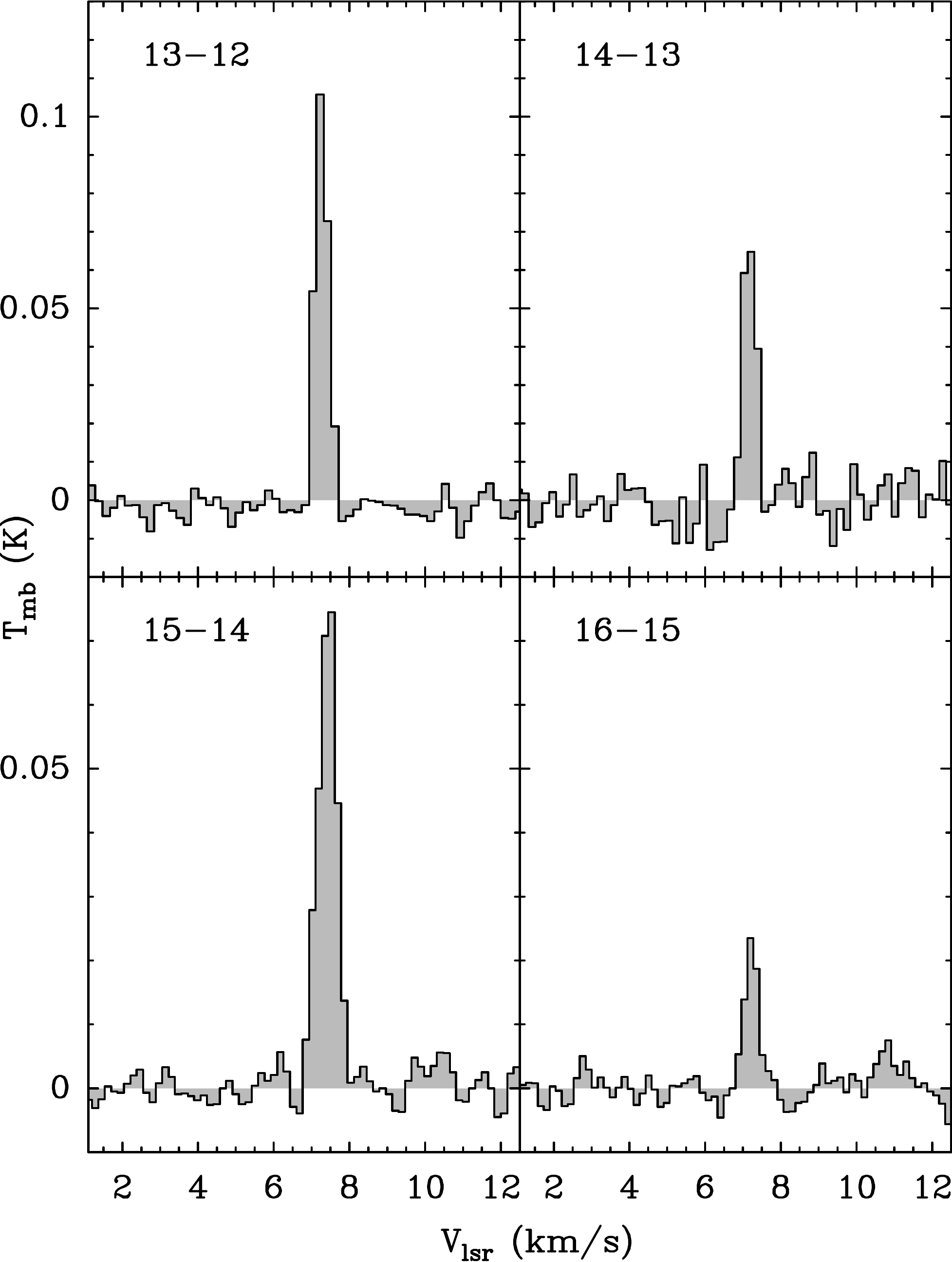}
   \caption{C$_3$S detected lines (in $\rm T_{mb}$).}
   \label{c3s}
 \end{figure}
 
  \begin{figure}
   \centering
   \includegraphics[width=7cm]{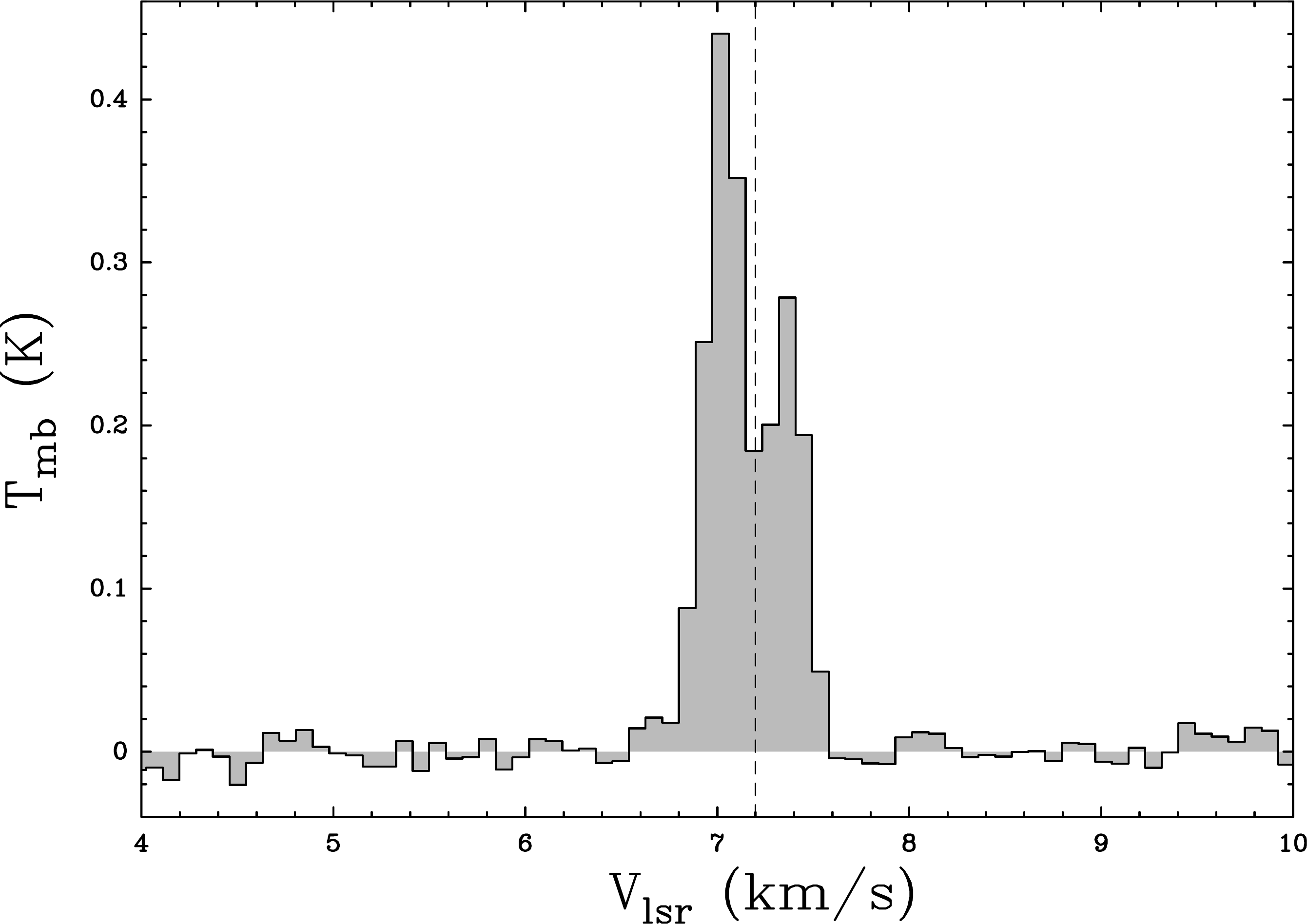}
   \caption{ortho-H$_2$S transition at 168.8 GHz (in $\rm T_{mb}$).}
   \label{h2s}
 \end{figure}

 \begin{figure*}
   \centering
   \includegraphics[width=10cm]{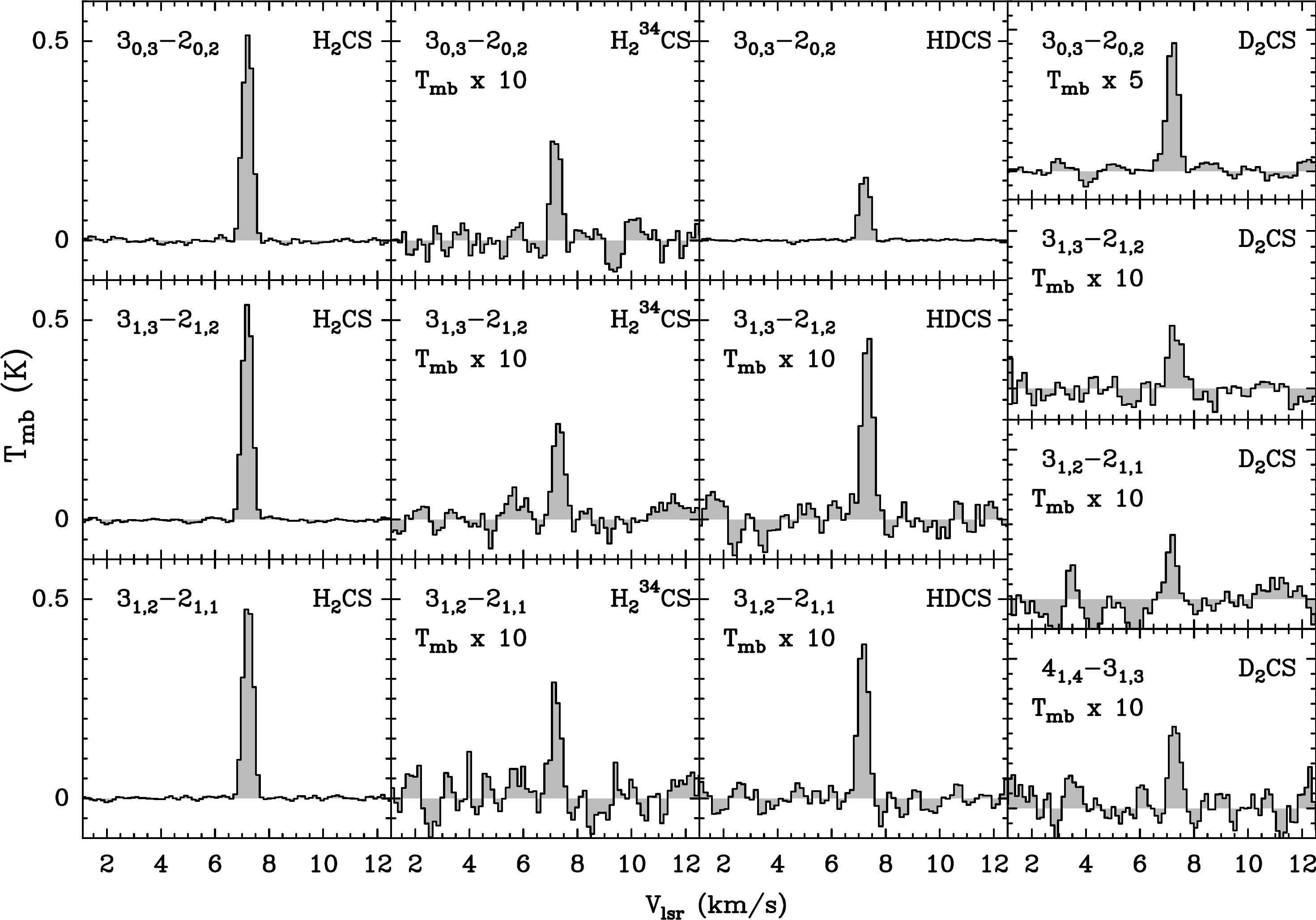}
   \caption{H$_2$CS, H$_2$$^{34}$CS, HDCS and D$_2$CS detected lines (in $\rm T_{mb}$).}
   \label{h2cs}
 \end{figure*}
 
  \begin{figure}
   \centering
   \includegraphics[width=7cm]{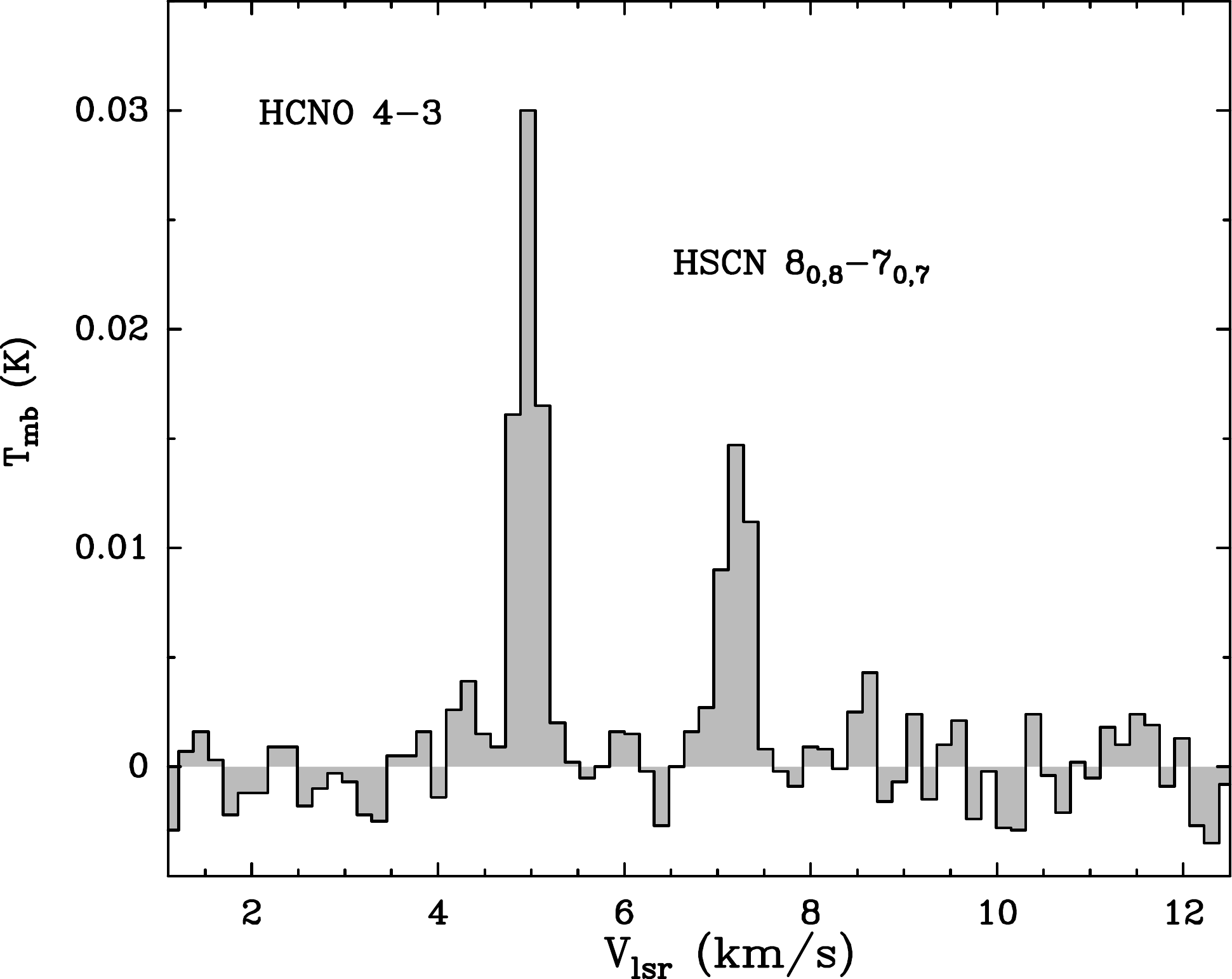}
   \caption{HSCN detected line (in $\rm T_{mb}$).}
   \label{hscn}
 \end{figure}
 
  \begin{figure}
   \centering
   \includegraphics[width=7cm]{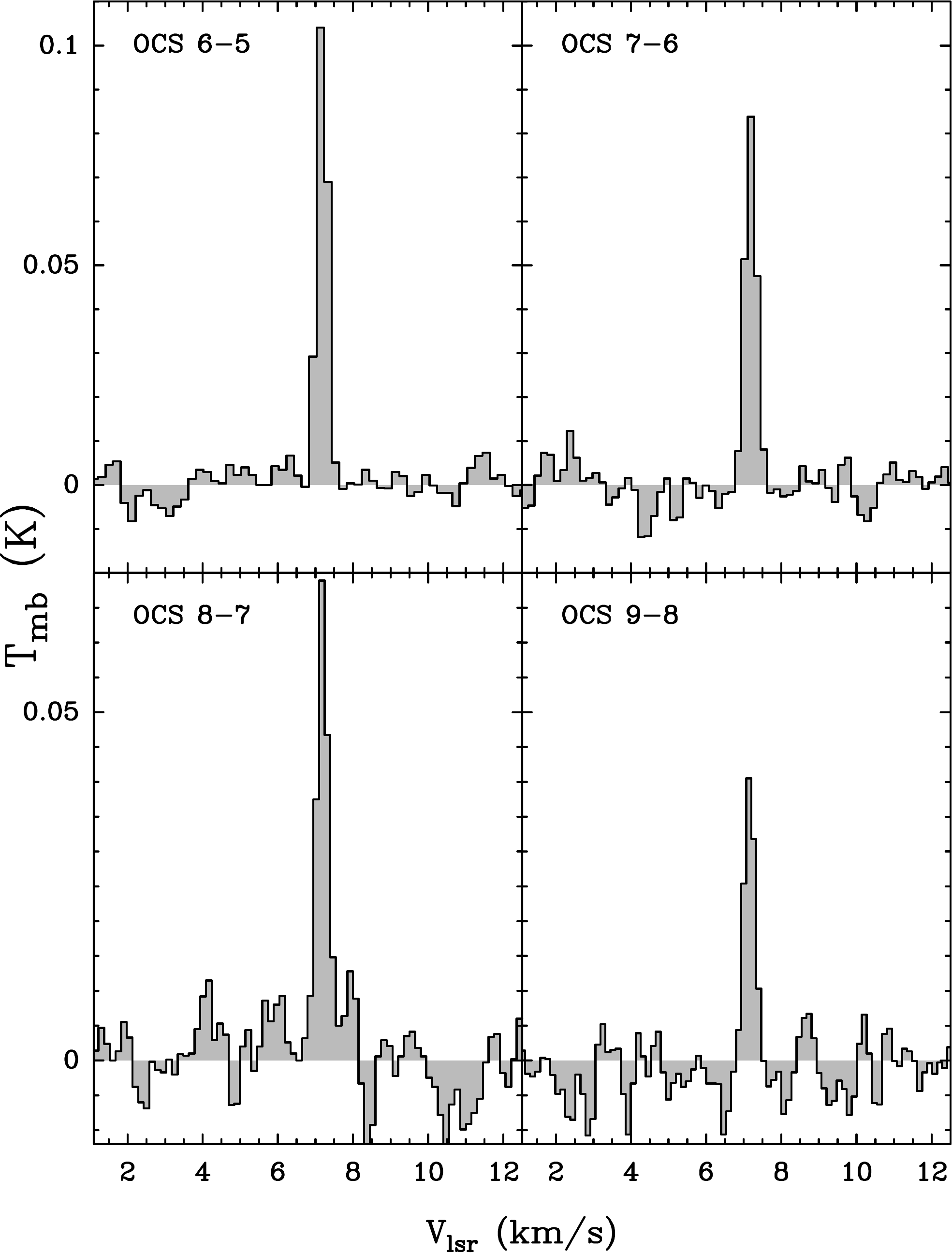}
   \caption{OCS detected lines (in $\rm T_{mb}$).}
   \label{ocs}
 \end{figure}

\begin{figure}
   \centering
   \includegraphics[width=7cm]{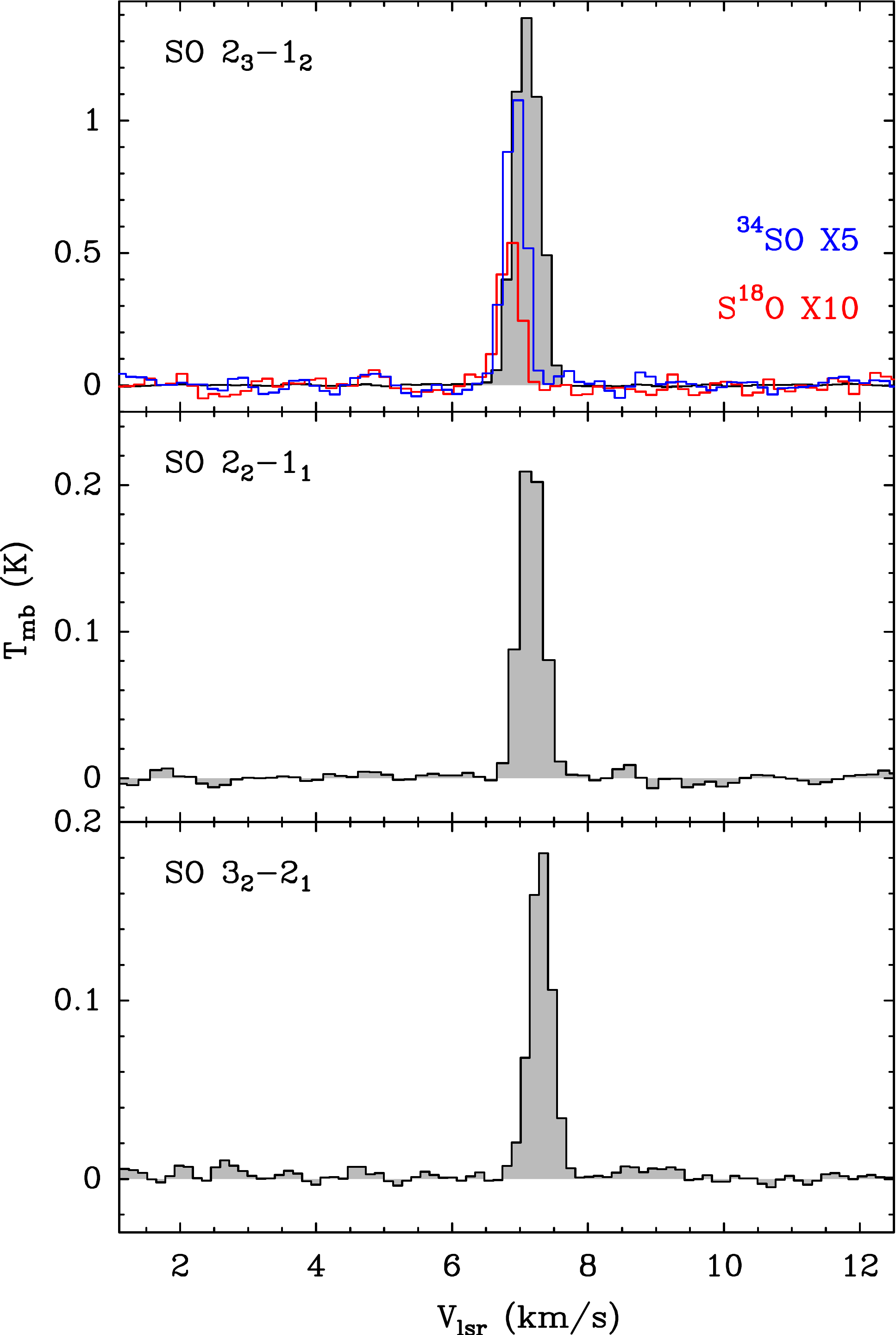}
   \caption{SO, S$^{18}$O and $^{34}$SO detected lines (in $\rm T_{mb}$).}
   \label{so}
 \end{figure}
 
 \begin{figure}
   \centering
   \includegraphics[width=7cm]{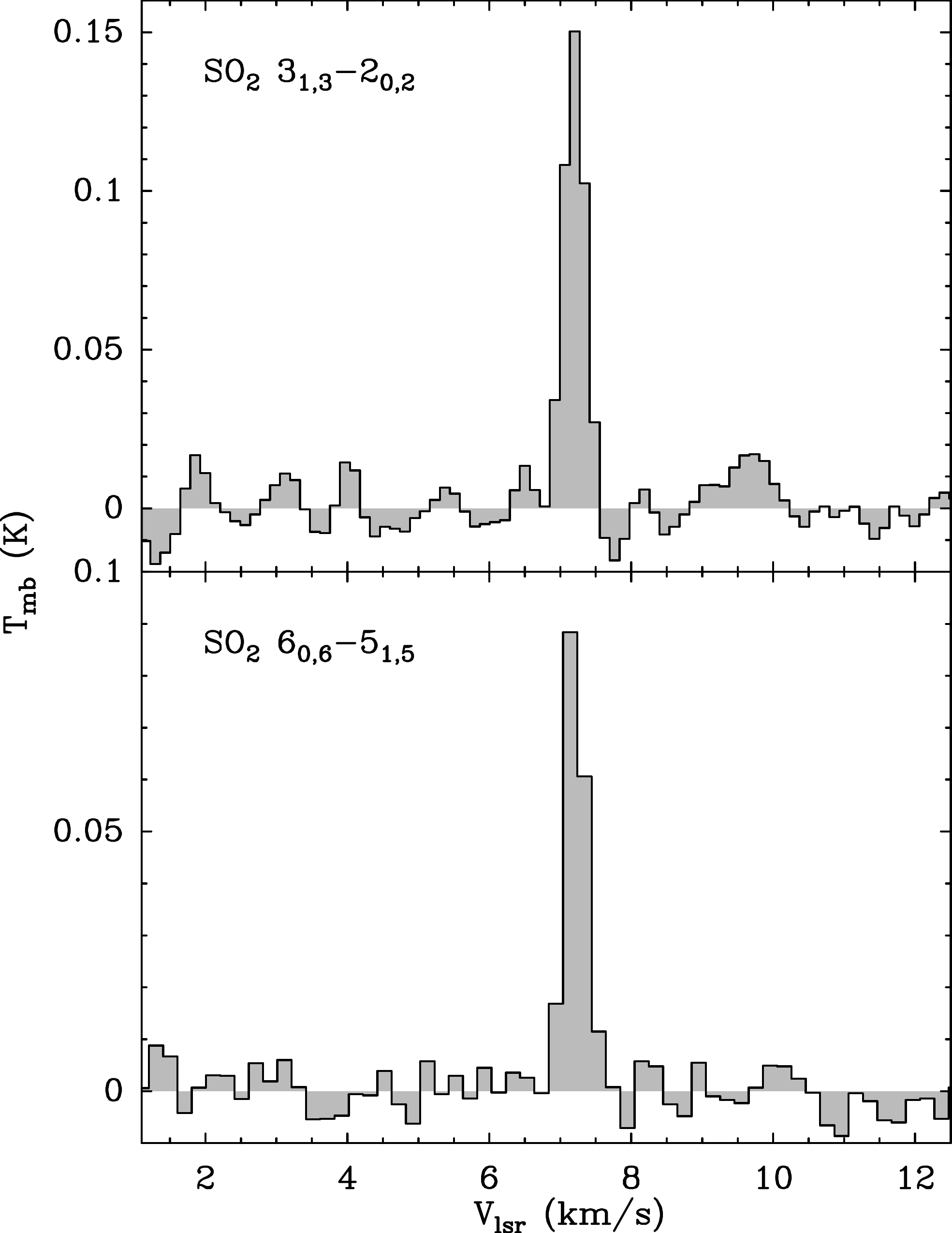}
   \caption{SO$_2$ detected lines (in $\rm T_{mb}$).}
   \label{so2}
 \end{figure}

 \begin{figure}
   \centering
   \includegraphics[width=7cm]{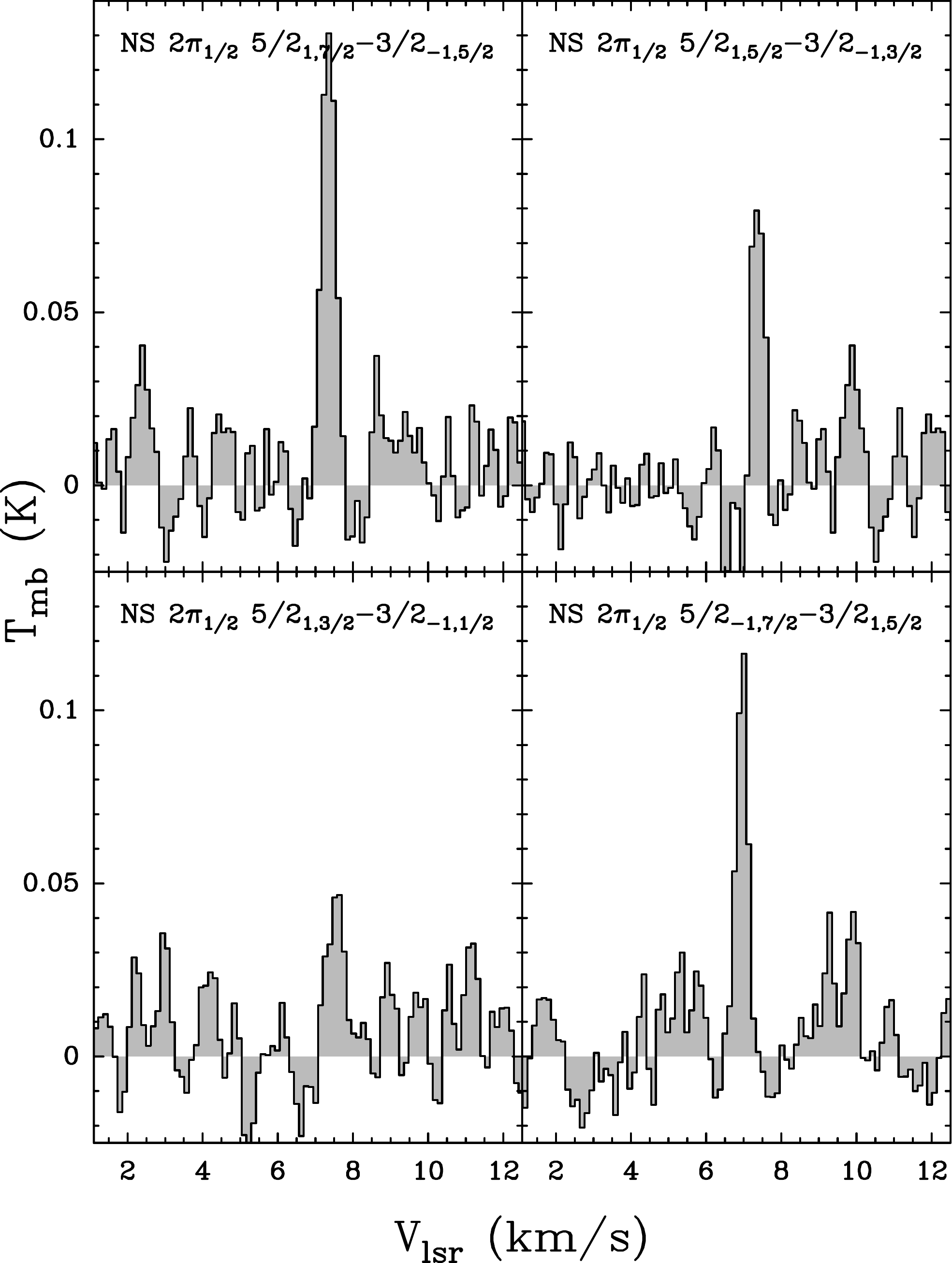}
   \caption{NS detected lines (in $\rm T_{mb}$).}
   \label{ns}
 \end{figure}

  \begin{figure}
   \centering
   \includegraphics[width=7cm]{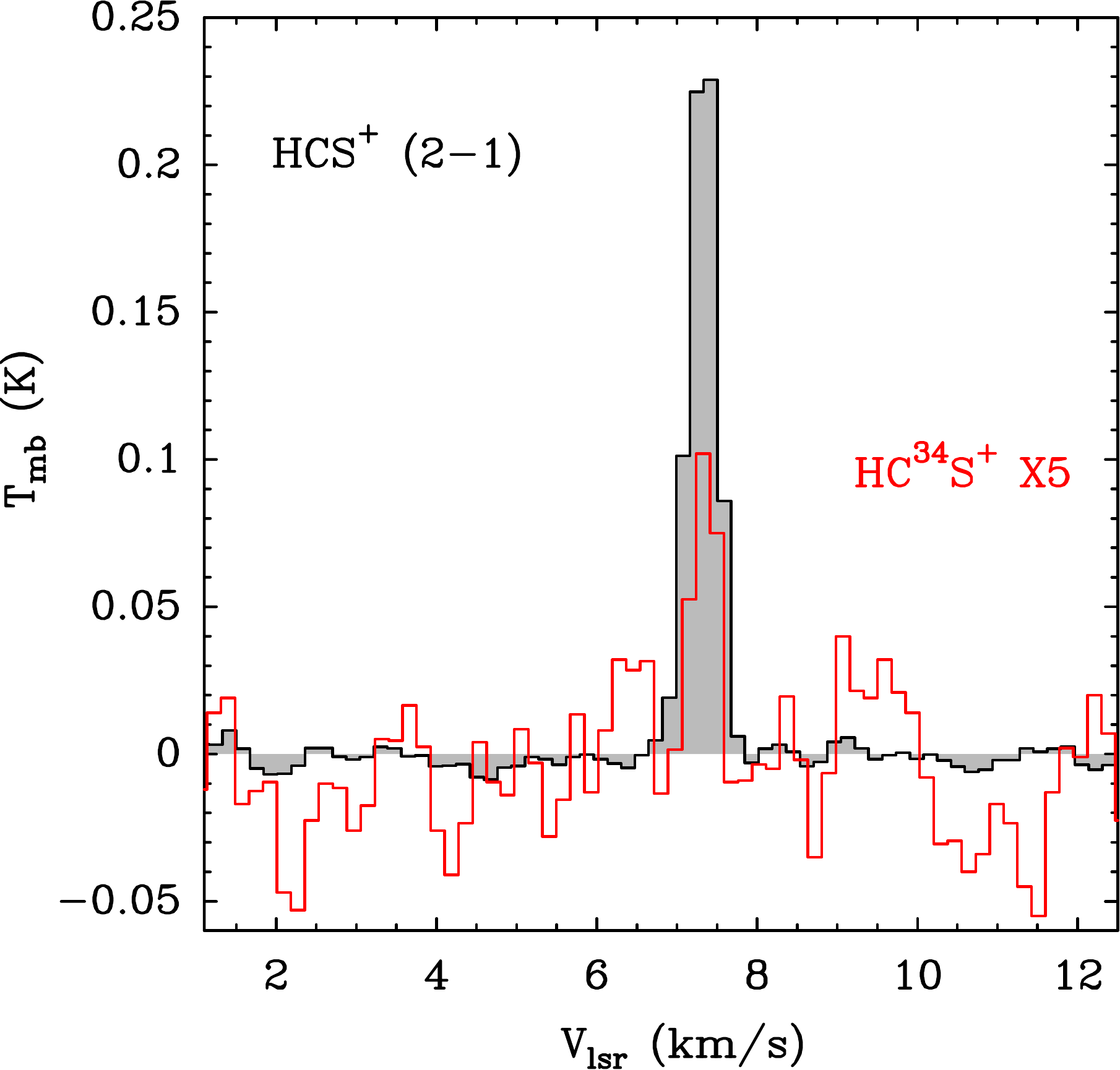}
   \caption{HCS$^+$ (black) and HC$^{34}$S (red) 2--1 detected line (in $\rm T_{mb}$).}
   \label{hcsp}
 \end{figure}
  
  \begin{figure}
   \centering
   \includegraphics[width=7cm]{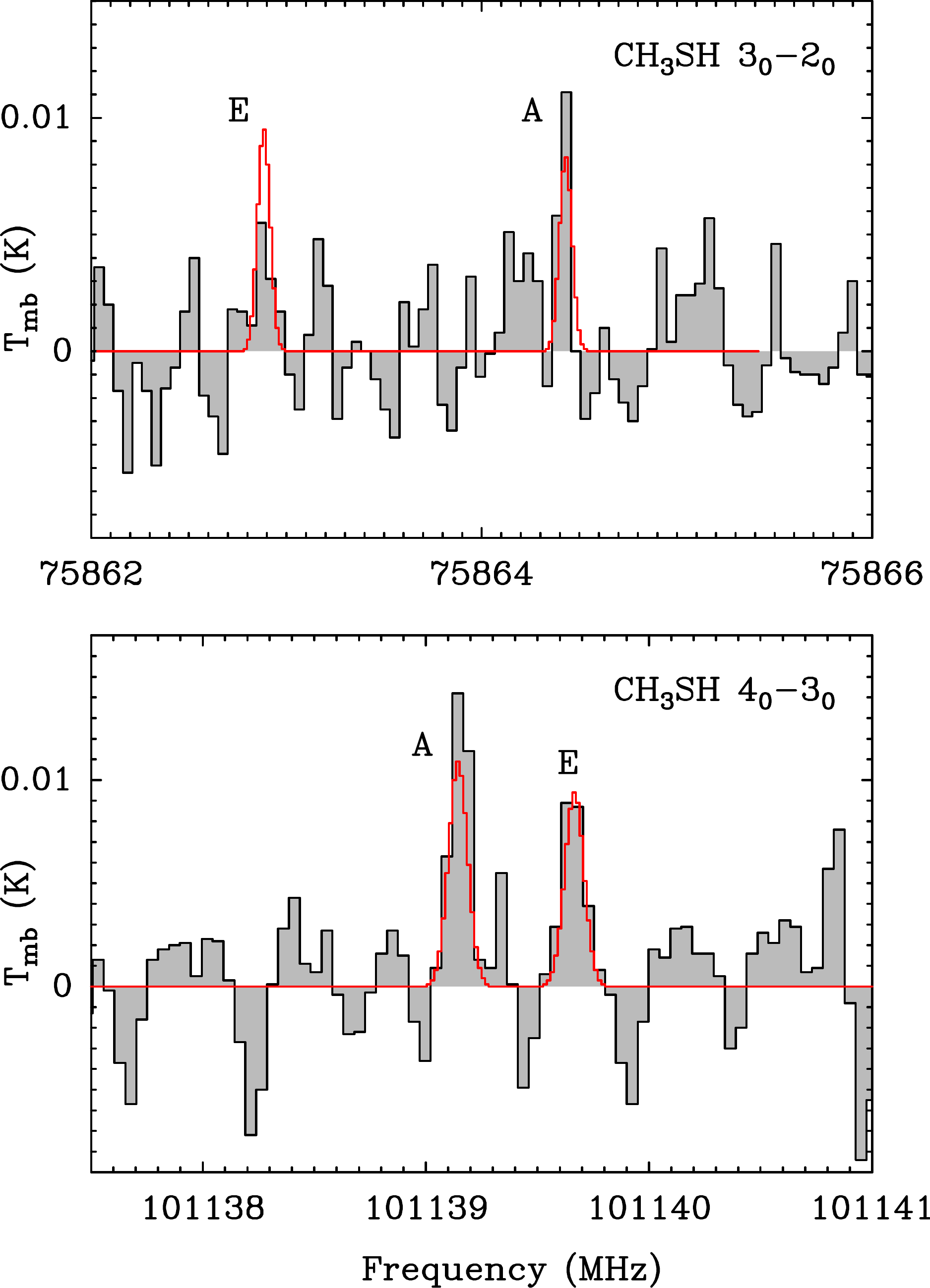}
   \caption{Tentative detection of CH$_3$SH. The LTE model ($\rm T_{ex}$ = 10K, N = 2.5 $\times$ 10$^{11}$ cm$^{-2}$, fwhm = 0.3 km/s) is over-plotted in red. The lines are in $\rm T_{mb}$.}
   \label{ch3sh}
 \end{figure}

\section{The puzzling behaviour of the H$_2$S line profile}

\begin{figure}
   \centering
   \includegraphics[width=\hsize]{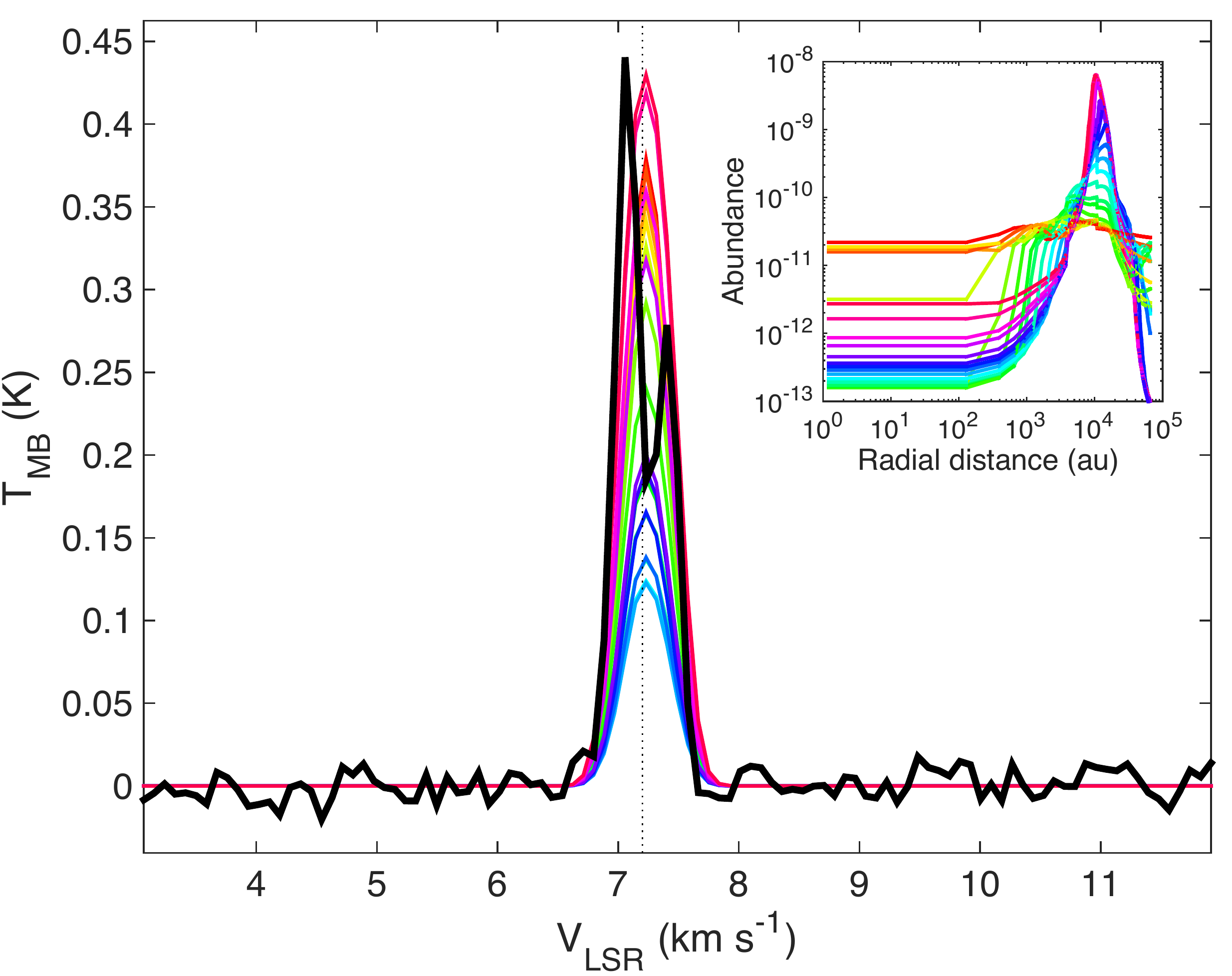}
   \caption{Variation of the calculated H$_2$S 1$_{1,0}$ -- 1$_{0,1}$ profile (in color) using the variation of the H$_2$S abundance from the \textsc{nautilus} modelling, for an age between 10$^6$ and 3~10$^6$ years, for models 1--3 (see Table \ref{init_dens}). The profiles are compared with the observed transition in black.}
   \label{H2S_spectra_nautilus}
 \end{figure}

From the radial variation of the H$_2$S abundance predicted by the {\sc nautilus} chemical modelling, we now are able to tentatively reproduce its puzzling line profile. We extracted the H$_2$S abundance as a function of radius for 20 ages between 10$^2$ and 10$^7$ years and used the {\sc gass}\footnote{Generator of Astrophysical Sources Structure} code \citep{quenard2017b} alongside the structure of L1544 to generate a 3D model of the source. We then use {\sc lime} \citep{brinch2010}, which is a 3D non-LTE ALI (Accelerated Lambda Iteration) continuum and gas line radiative transfer treatment, to produce the line intensity data cube. Since the H$_2$S collision coefficients are not available, we instead used the ortho-H$_2$O collision coefficients with para--H$_2$ from \citet{dubernet2006} as a first approximation. The \textsc{lime} data cubes have been post-processed by {\sc gass} to extract the H$_2$S spectra and compare them to the observation. We found that no model extracted from {\sc nautilus} can satisfactorily reproduce the absorption feature seen at the V$_{\rm LSR}$ of L1544 in Fig. \ref{h2s}. Fig. \ref{H2S_spectra_nautilus} shows the result from the {\sc nautilus} chemical modelling for an age between 10$^6$ and 3~10$^6$ years, for models 1-3 (see Table \ref{init_dens}).

\begin{figure}
   \centering
   \includegraphics[width=\hsize]{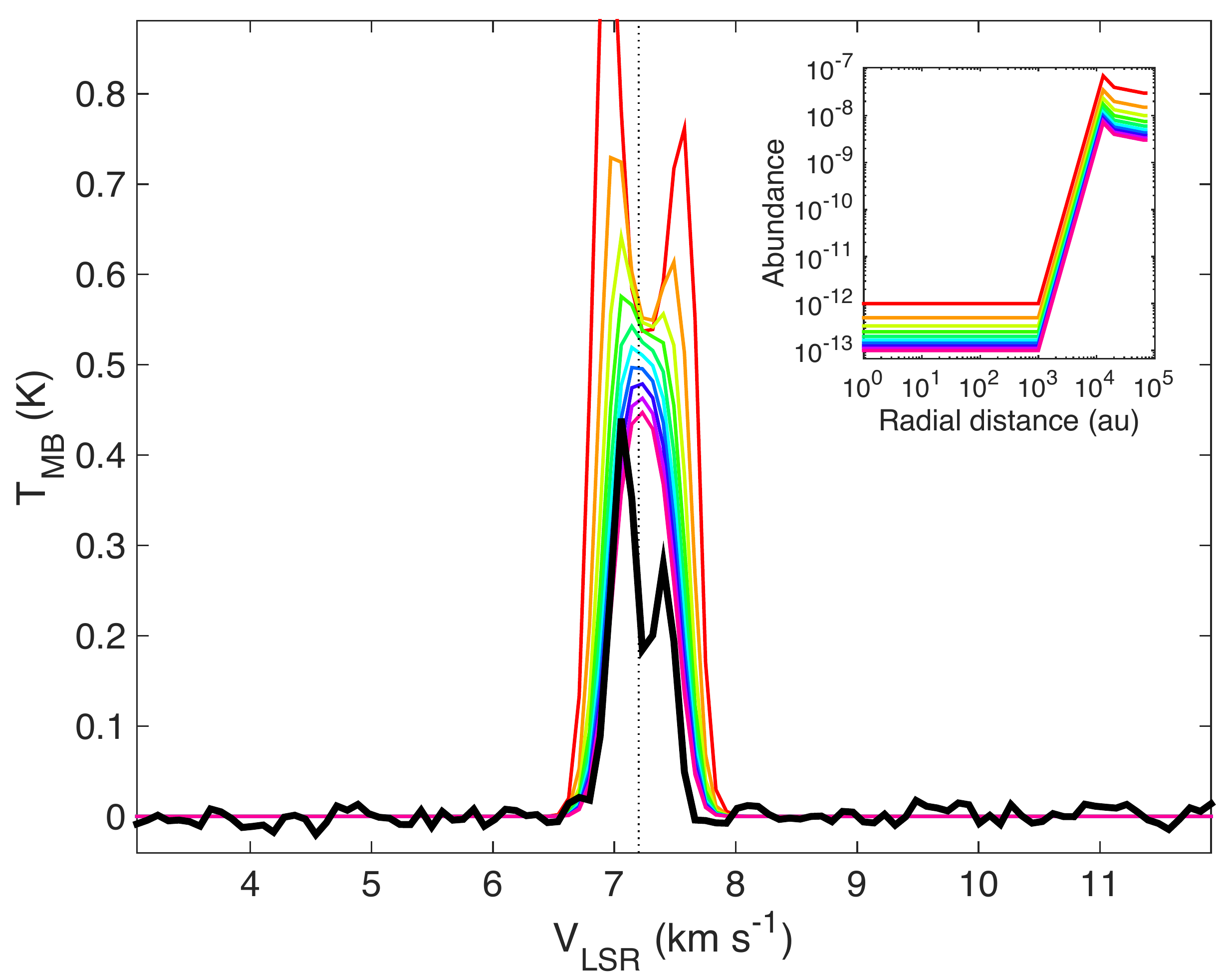}
   \caption{Variation of the calculated H$_2$S 1$_{1,0}$ -- 1$_{0,1}$ profile (in color) using various {\it ad-hoc} abundance profiles (see the inset), compared with the observation (in black).}
   \label{H2S_spectra}
 \end{figure}

Considering the similarities between the H$_2$O and H$_2$S molecules, we then tried the same water abundance profile as the one found by \citet{quenard2016}. We multiplied this profile by factors ranging [1--10] to take into account the different abundances of these two species. The resulting line shapes are shown in Fig. \ref{H2S_spectra} with a small inset presenting some of the abundance profiles used. As one can notice, the absorption can be reproduced for models with a similar (or close by a factor of $\sim$3) abundance profile than the one derived for water by \citet{quenard2016}. However the line is too intense by a scaling factor of $\sim$2.5 (see e.g. red and orange line profiles of Fig. \ref{H2S_spectra}). This could be caused by the collision coefficient that we have used as an approximation, which might not be appropriate for H$_2$S. \\ 
To consolidate our findings, we have also considered different {\it ad-hoc} abundance profiles to try to better reproduce the observed absorption feature. We have tried $\sim$100 different abundance profile shapes (such as constant abundance as a function of the radius, slope of abundances between two radii, step-function of abundances starting at different radii) but none of them produces a better fit to the observation. Dilution has not been considered since it contradicts the results from the chemical modelling and the maps performed in this source by \citet{spezzano2017}. We therefore concluded that more accurate collision rates might be needed for H$_2$S with at least para--H$_2$ and their use will possibly help to better reproduce the observed line profile and constrain the chemistry.\\

\section{Additional chemical modelling}

 \begin{figure*}
   \centering
   \includegraphics[width=\hsize]{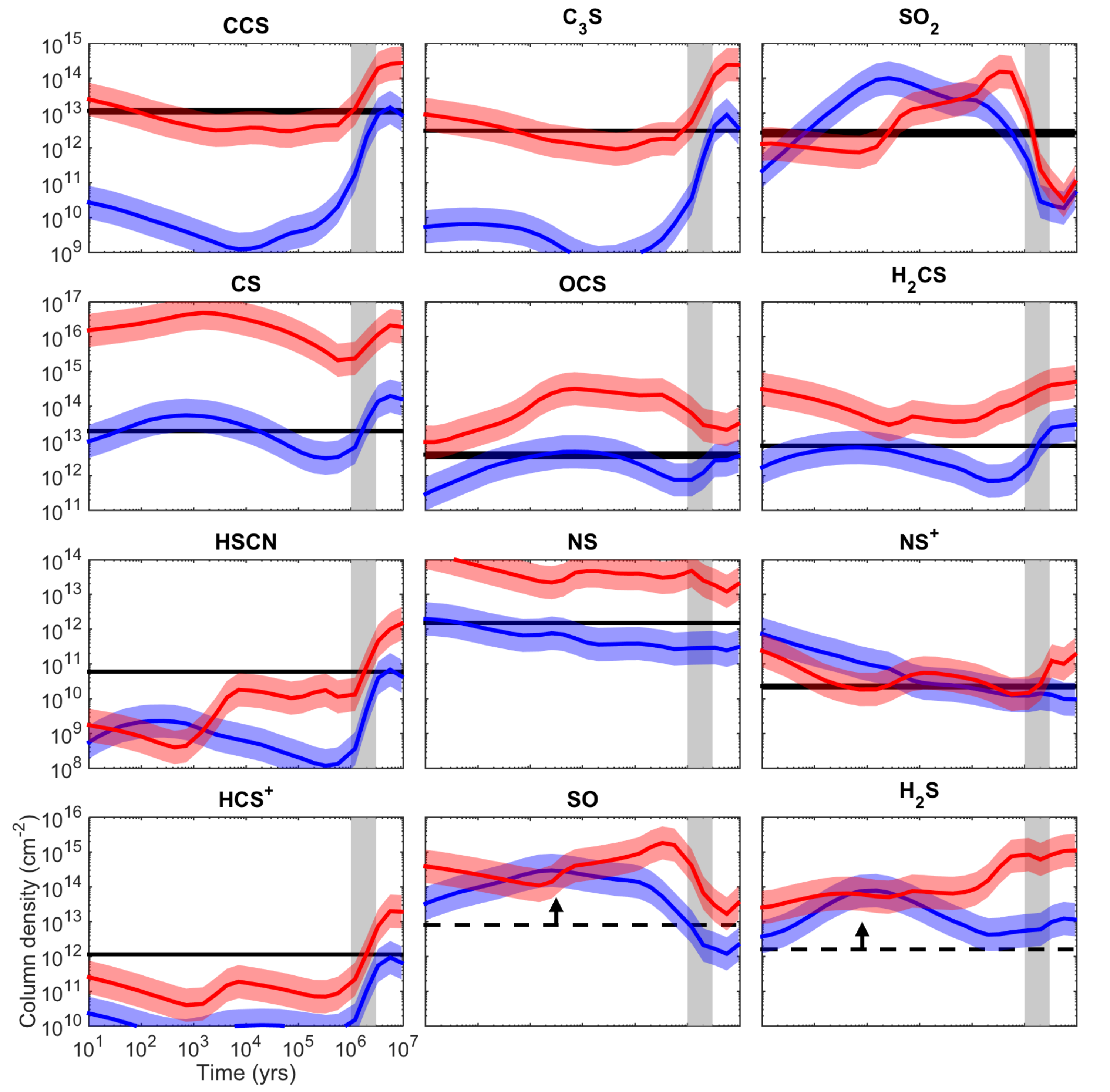}
   \caption{Variation of the modelled column density as a function of time and the comparison with the observed column density (black horizontal line). The blue lines correspond to model 2 (sulphur depletion: S/H=$8~10^{-8}$) and the red ones correspond to model 5 (sulphur non depletion: S/H=$1.5~10^{-5}$) as shown in Table \ref{init_dens}. The thickness of the black line corresponds to the error bar of the observed column densities. A variation by a factor of three of modelled column densities is shown in corresponding coloured areas. The dashed black horizontal line for SO and H$_2$S correspond to the lower limit on the computation of the total column density. The gray vertical area highlights an age between [1--3] $\times$ 10$^6$ years (see text).}
   \label{Ncol_density_3e3}
 \end{figure*}
 
  \begin{figure*}
   \centering
   \includegraphics[width=\hsize]{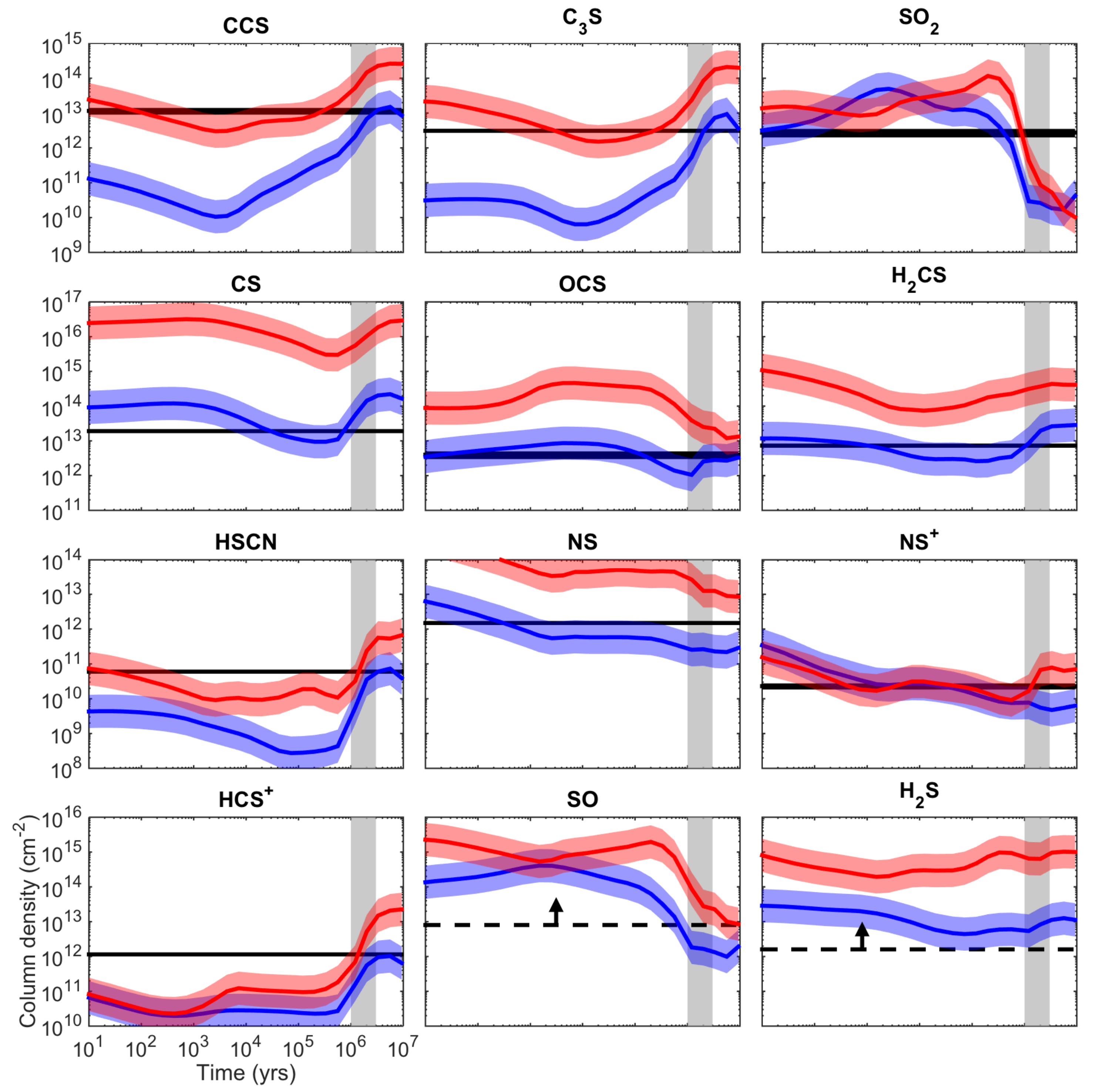}
   \caption{Variation of the modelled column density as a function of time and the comparison with the observed column density (black horizontal line). The blue lines correspond to model 3 (sulphur depletion: S/H=$8~10^{-8}$) and the red ones correspond to model 6 (sulphur non depletion: S/H=$1.5~10^{-5}$) as shown in Table \ref{init_dens}. The thickness of the black line corresponds to the error bar of the observed column densities. A variation by a factor of three of modelled column densities is shown in corresponding coloured areas. The dashed black horizontal line for SO and H$_2$S correspond to the lower limit on the computation of the total column density. The gray vertical area highlights an age between [1--3] $\times$ 10$^6$ years (see text).}
   \label{Ncol_density_2e4}
 \end{figure*}


\bsp	
\label{lastpage}
\end{document}